\documentclass[preprint,showpacs,preprintnumbers,amsmath,amssymb]{revtex4}
\usepackage{amsmath,amssymb,graphics,epsfig,subfigure,color}
\usepackage{color}

\begin{document}
\newcommand {\nn}    {\nonumber}
\renewcommand{\baselinestretch}{1.3}

\title{New localization mechanism and Hodge duality for $q-$form field}
\author{Chun-E Fu$^a$\footnote{fuche08@lzu.edu.cn, corresponding author},
        Yu-Xiao Liu$^b$\footnote{liuyx@lzu.edu.cn},
        Heng Guo$^c$\footnote{guoh06@lzu.edu.cn},
        Sheng-Li Zhang$^a$\footnote{zhangsl@mail.xjtu.edu.cn}}.
 \affiliation{
$^a$School of Science, Xi'an Jiaotong University, Xi'an 710049, P. R. China\\
$^b$Institute of Theoretical Physics,
Lanzhou University, Lanzhou 730000, P. R. China\\
$^c$School of Physics and Optoelectronic Engineering,
Xidian University, Xi'an 710071, P. R. China}


\begin{abstract}
{
  In this paper, we investigate the problem of localization and {the} Hodge duality for a $q-$form field on a $p-$brane with codimension one. By a general Kaluza-Klein (KK) decomposition without gauge fixing, we obtain two Schr\"{o}dinger-like equations for {two types of KK modes} of the bulk $q-$form field, which determine the localization and mass spectra of these KK modes. It is found that there are two types of zero modes (the $0-$level modes): a $q-$form zero mode and a $(q-1)-$form one, which cannot be localized on the brane at the same time.
  For the $n-$level KK modes, there are two interacting KK modes, a massive $q-$form KK mode and a massless $(q-1)-$form one.
  By analyzing {gauge invariance of the effective action} and choosing a gauge condition, the $n-$level massive $q-$form KK mode {decouples from} the $n-$level massless $(q-1)-$form one. It is {also} found that the Hodge duality in the bulk naturally becomes two dualities on the brane. The first one is the Hodge duality between a $q-$form zero mode and a $(p-q-1)-$form one,  or between a $(q-1)-$form zero mode and a $(p-q)-$form one. The second duality is between two group KK modes: one is an $n-$level massive $q-$form KK mode with mass $m_n$ and an $n-$level massless $(q-1)-$form mode; another is an $n-$level $(p-q)-$form one with the same mass $m_n$ and an $n-$level massless $(p-q-1)-$form mode. Because of the dualities, the effective field theories on the brane for the KK modes of the two dual bulk form fields are physically equivalent.
  }
  \end{abstract}

\pacs{04.50.-h, 11.27.+d }

\maketitle

\section{Introduction}

When the Arkani-Hamed-Dimopoulos-Dvali (ADD) \cite{Antoniadis1998} and {the} Randall-Sundrum (RS) \cite{Randall1999a,Randall1999b} brane-world models were brought up, they opened a new avenue to solve the long-standing hierarchy problem and the cosmology problem \cite{Rubakov2,ArkaniHamed1998rs,Cosmological2000,PhysRevLett.86.4223,coscon2009,Neupane2010ey,
BraneFR2012,ThickBrane20093}. Since then the brane-world and extra dimension theories have received more and more attention \cite{ThickBraneDewolfe,ThickBrane20093,Gregory:2000jc,Kaloper2000jb,PhysRevD.66.024024,ThickBrane2002,
ThickBrane2003,Blochbrane2004,ThickBraneBazeia2006,ThickBrane20071,ThickBraneDzhunshaliev2008,
KeYang2009Weyl,ThickBraneZhongYuan2011JHEP,Liu:2012rc,
Liu:2012mia,BraneFR2012,Bazeia:2013uva,Das:2014tva}.

In the brane-world theory, one of the most important and interesting subjects is to investigate the Kaluza-Klein (KK) modes of various fields \cite{Grossman:1999ra,Bajc:1999mh,Gremm:1999pj,Chang1999nh,RandjbarDaemi:2000cr,Kehagias:2000au,
Duff2000se,Oda2001,Ringeval:2001cq,Ichinose:2002kg,Koley:2004at,Davies:2007xr,
Liu2008WeylPT,Archer2011,Jones:2013ofa,Xie2013rka,Cembranos:2013qja,Costa:2013eua,Zhao2014gka,
Carrillo-Gonzalez:2014dka,Kulaxizi:2014yxa}, which are the codes of the extra dimensions. In this work we are interested in the KK modes of a bulk massless $q-$form field on {a} $p-$brane world with codimension one. It is known that the $0-$form and $1-$form fields are the scalar and vector fields, respectively, and the usual $2-$form field is the Kalb-Ramond field, which is used to describe the torsion of space-time in Einstein-Cartan theory. The higher-form ($q>2$) fields are new types of particles in a higher-dimensional space-time with dimension $D>4$, which are useful for some unknown problems such as the cosmological constant problem or dark energy problem \cite{PhysRevD.88.123512,PhysRevLett.86.4223}.

The $p-$brane considered here has $p$ spacial dimensions, and is embedded in a $D=p+2$ dimensional space-time with one extra dimension perpendicular to the brane. Although the realistic world is the $3-$brane, the higher dimensional branes with $p>3$ ($D>5$) may also have realistic applications if the branes have three infinite large dimensions (which are those we can feel) and $p-3$ finite size dimensions with topology $S^1 \times S^1 \times ... \times S^1 = T^{p-3}$ and small enough radius. It is also possible that there are more extra dimensions {outside} the $p-$brane. In this work we only consider the simplest case. The line element of the space-time is assumed as a RS-like one,
\begin{eqnarray}
  ds^2=\text{e}^{2A(z)}(\hat{g}_{\mu\nu}(x^{\lambda}) dx^\mu\ dx^\nu+dz^2),\label{metric}
\end{eqnarray}
where $A(z)$ is the warp factor that is only the function of $z$, and $\hat{g}_{\mu\nu}(x^{\lambda})=\eta_{\mu\nu}$ is the induced flat metric on the brane. Moreover we regard the bulk $q-$form field as a small perturbation around the $p-$brane and neglect its backreaction on the background geometry.

To investigate the KK modes of a bulk field, we should propose a localization mechanism for the field. Some work has been carried out on the localization of the $q-$form field \cite{Duff2000se,PhysRevD.32.3238,PhysRevD.37.1690,PRLKR2002,QformRS,qRSdilation-eprintv2,
QformThickRS,Youm:2001dt,Fu2012sa,Fu:2013ita,Ho:2013paa,Hinterbichler:2013kwa,Jardim2014vba,Ho2014una,Smailagic2000hr}, where the authors usually first chose a gauge to make the localization mechanism simpler. But these gauge choices make us only see parts of the whole localization information. In this paper, we will try to find a new localization mechanism for the $q-$form field without any gauge choice in order to give a whole view of the field's localization.

To this end, we first give a general KK decomposition for a bulk $q-$form field $X_{M_1M_2\cdots M_q}$ without any gauge choice:
\begin{subequations}\label{KKdecomposition}
\begin{eqnarray}
  X_{\mu_1\cdots \mu_q}(x_{\mu},z)&=&\sum_{n}\hat{X}_{\mu_1\cdots \mu_q}^{(n)}(x^{\mu})\;U_1^{(n)}(z)\text{e}^{a_1A(z)}, \label{kk1} \\
  X_{\mu_1\cdots \mu_{q-1}z}(x_{\mu},z)&=&\sum_{n}\hat{X}_{\mu_1\cdots \mu_{q-1}}^{(n)}(x^{\mu})\;U_2^{(n)}(z)\text{e}^{a_2A(z)},\label{kk2}
\end{eqnarray}
\end{subequations}
where the $\hat{}$ denotes the effective quantities on the brane, the index $n$ marks different KK modes, $U_{i}(z)$ are only the function of the extra dimension coordinate $z$, and $a_i$ are constants. Here we have classified the bulk $q-$form field into two types, i.e., $X_{\mu_1\cdots \mu_{q-1}z}$ containing the index of $z$ and $X_{\mu_1\cdots \mu_q}$ not containing, because they have different effective fields on the brane.
The effective fields on the brane for $X_{\mu_1\cdots \mu_{q-1}z}$ and $X_{\mu_1\cdots \mu_q}$ are the $(q-1)-$form and $q-$form fields, respectively.

Then through the dimensional reduction we will get the effective action for the $q-$form field, the orthonormality conditions for the KK modes, and the equations of motion of $U_1^{(n)}(z)$ and $U_2^{(n)}(z)$, which are found to be two Schr\"{o}dinger-like equations. With the orthonormality conditions and the two Schr\"{o}dinger-like equations we can get the mass spectra of the KK modes and analyze their characters in any $p-$brane world model.

It will be finally found that for any $q-$form field there are two types of {massless $0-$level} KK modes, a $q-$form mode and a $(q-1)-$form one, which cannot be localized on the $p-$brane at the same time.  {There are also two interacting $n-$level KK modes, i.e., a massive $q-$form KK mode and a massless $(q-1)-$form one.}

With the general KK decomposition, we also will find that the Hodge duality in the bulk naturally turns out to be two dualities, i.e., the Hodge duality on the brane between a massless $q-$form mode and a $(p-q-1)-$form one, and  a new duality between two group KK modes: one is an $n-$level massive $q-$form KK mode with mass $m_n$ and an $n-$level massless $(q-1)-$form mode; another is an $n-$level $(p-q)-$form one with the same mass $m_n$ and an $n-$level massless $(p-q-1)-$form mode. Because of the two dualities, there are some interesting phenomena. For example, as a massless vector ($1-$form) field on a $3-$brane is dual to itself according to the Hodge duality on the brane, we cannot know it is from a bulk $1-$form or a $2-$form field.

This paper is organized as follows. We first investigate the new localization mechanism in Sec.~\ref{localizationmechanism}, and then discuss the massless and bound massive KK modes, respectively, in Secs. \ref{massless} and \ref{massive}. {In Sec. \ref{dualitymassive} we show the new duality derived from the Hodge duality in the bulk.} Finally, we give a brief conclusion in Sec. \ref{conclusion}.

\section{A new localization mechanism and Hodge duality}\label{localizationmechanism}

In a brane-world background, there are usually four steps to investigate the localization of a bulk $q-$form field, for which the action is \cite{Duff2000se}
\begin{eqnarray}\label{bulkaction}
  S&=&{-\frac{1}{2(q+1)!}}\int d^D x\sqrt{-g}\;Y^{M_1M_2\cdots M_{q+1}}Y_{M_1M_2\cdots M_{q+1}},
\end{eqnarray}
{ with $Y_{M_1M_2...M_{q+1}}$ the field strength defined as $Y_{M_1M_2...M_{q+1}}=\partial_{[M_1}X_{M_2...M_{q+1}]}$}.
\begin{itemize}
  \item{Firstly, choose a simple gauge for the $q-$form field, such as
  \begin{equation}\label{gauge1}
   {X_{\mu_1\cdots \mu_{q-1}z}(x^{\mu},z)=0,}
  \end{equation}
   and make a KK decomposition for other components of the field:
   \begin{equation}
    X_{\mu_1\cdots \mu_q}(x^{\mu},z)=\sum_n\hat{X}_{\mu_1\cdots \mu_q}^{(n)}(x^{\mu})\;U_1^{(n)}(z)\;\text{e}^{a_1A}.  \label{KKdecompositionOld}
   \end{equation}
   }
  \item{Secondly, substitute the KK decomposition into the equations of motion for the $q-$form field:
   \begin{subequations}\label{equ}
   \begin{eqnarray}
   \partial_{\mu_1} (\sqrt{-g}\;Y^{\mu_1\mu_2...\mu_{q+1}})
   +{\partial_z(\sqrt{-g}\;Y^{z\mu_2...\mu_{q+1}})}&=& 0, \label{equ1}\\
   \partial_{\mu_1} (\sqrt{-g}\;Y^{\mu_1\mu_2...\mu_qz})&=& 0,\label{equ2}
   \end{eqnarray}
   \end{subequations}
   and use the gauge choice (\ref{gauge1}),
   then obtain a Schr\"{o}dinger-like equation for the KK modes $U_1^{(n)}(z)$. }
  \item{Thirdly, with the KK decomposition into the action of the bulk field (\ref{bulkaction}), and assuming the orthonormality conditions for localizing the KK modes on the brane, the effective action for the KK modes will be found.}
  \item{ Lastly, with a background solution, by solving the equation of $U_1^{(n)}(z)$, the mass spectra of the KK modes are obtained, and their characters can also be analyzed.}
\end{itemize}
The reason for choosing the gauge (\ref{gauge1}) is to obtain a simpler equation of $U_1^{(n)}(z)$ from (\ref{equ1}), which is used to find the {mass spectra} and the wave functions of the KK modes satisfying the orthonormality conditions.

If we substitute the general KK decompositions (\ref{KKdecomposition}) into the field equations (\ref{equ}), we will get complex field equations for the KK modes $U_1^{(n)}(z)$ and $U_2^{(n)}(z)$. However, we indeed need these equations to discuss the localization of the $q-$form field.

In order to investigate this problem, we will compare the equations of motion for the KK modes $U_1^{(n)}(z)$ and $U_2^{(n)}(z)$ derived in two ways. One is from the effective action, which is obtained by the KK reduction for the fundamental action of the $q-$form field. Another is from (\ref{equ}) as well as the KK decomposition (\ref{KKdecomposition}). Let us show the details.

We first would like to get the effective action for the $q-$form field. With the KK decomposition (\ref{KKdecomposition}), the corresponding components of the field strength become
\begin{eqnarray}
  Y^{\mu_1\cdots \mu_{q+1}}&=&
    \sum_n \hat{Y}^{\mu_1\cdots \mu_{q+1}}_{(n)}(x^{\mu}) \;U_1^{(n)}(z)\text{e}^{(a_1-2(q+1))A},\\
  Y^{\mu_1\cdots \mu_{q}z}   &=&
    {\frac{q}{q+1}\sum_n\hat{Y}^{\mu_1\cdots \mu_q}_{(n)}(x^{\mu})}\;U_2^{(n)}(z)\text{e}^{(a_2-2(q+1))A}\nonumber\\
  &&+\;{\frac{1}{q+1}
    \sum_n\hat{X}_{(n)}^{\mu_1\cdots \mu_q}}(x^{\mu})\;
\partial_z \bigg(U_1^{(n)}(z)\text{e}^{a_1A}\bigg)\text{e}^{-2(q+1)A},\label{Y2}
\end{eqnarray}
where the indices of the quantities with $\hat{~}$ are raised or lowered by the {induced} metric $\hat{g}^{\mu\nu}(x)$ or $\hat{g}_{\mu\nu}(x)$. Substituting the above expressions into the action for the $q-$form field, we have:
\begin{eqnarray}\label{actioneffective}
  S_q
  &=&{-\frac{1}{2(q+1)!}}\int d^D x\sqrt{-g}
      \bigg(Y^{\mu_1\cdots \mu_{q+1}}Y_{\mu_1\cdots \mu_{q+1}}
            +{(q+1)}Y^{\mu_1\cdots \mu_qz}Y_{\mu_1\cdots \mu_qz}\bigg),\nonumber\\
   &=&{-\frac{1}{2(q+1)!}}
     \sum_n\sum_{n'}\bigg[ I^{(1)}_{nn'}\int d^{p+1}x \sqrt{-\hat{g}} \;
           \hat{Y}_{(n)}^{\mu_1\cdots \mu_{q+1}}
           \hat{Y}^{(n')}_{\mu_1\cdots \mu_{q+1}} \nonumber\\
   &&\quad\quad\quad+ I^{(2)}_{nn'}\int d^{p+1}x \sqrt{-\hat{g}}\;
           \hat{Y}_{(n)}^{\mu_1\cdots \mu_{q}}
           \hat{Y}^{(n')}_{\mu_1\cdots \mu_{q}}\nonumber\\
   &&\quad\quad\quad+ I^{(3)}_{nn'}\int d^{p+1}x \sqrt{-\hat{g}}\;
           \hat{X}_{(n)}^{\mu_1\cdots \mu_{q}}
           \hat{X}^{(n')}_{\mu_1\cdots \mu_{q}}\nonumber\\
   &&\quad\quad\quad+  2I^{(4)}_{nn'}\int d ^{p+1}x\sqrt{-\hat{g}}\;
           \hat{Y}_{(n)}^{\mu_1\cdots \mu_{q}}
           \hat{X}^{(n')}_{\mu_1\cdots \mu_{q}}\bigg],
\end{eqnarray}
where we have let $a_1=a_2=(2q-p)/2$, and supposed that $U_1^{(n)}(z)$ and $U_2^{(n)}(z)$ satisfy the following orthonormality conditions:
\begin{subequations}\label{OrthonormalityCondition}
\begin{eqnarray}
  I^{(1)}_{nn'}&\equiv&\int dz\;U_1^{(n)}U_1^{(n')}
    =\delta_{nn'},\label{condition1}\\
  I^{(2)}_{nn'}&\equiv&\frac{q^2}{q+1}\int dz\; U_2^{(n)}U_2^{(n')}
    ={(q+1)\;}\delta_{nn'},\label{condition2}
\end{eqnarray}
\end{subequations}
and $I^{(3)}_{nn'}$ and $I^{(4)}_{nn'}$ are given by
\begin{eqnarray}
    I^{(3)}_{nn'} &\equiv& \frac{1}{q+1}\int dz\; \text{e}^{(p-2q)A}
        \partial_z(U^{(n)}_1\text{e}^{a_1A})
        \partial_z(U^{(n')}_1\text{e}^{a_1A})
   , \label{condition3}\\
  I^{(4)}_{nn'}    &\equiv&  \frac{q}{q+1}\int dz\; \text{e}^{(p-2q)A/2}
        U^{(n)}_2\;\partial_z(U^{(n')}_1\text{e}^{a_1A}).  \label{condition4}
\end{eqnarray}

For the effective action (\ref{actioneffective}), it is necessary to analyze the mass dimensions of the constants $I^{(3)}_{nn'}$ and $I^{(4)}_{nn'}$ in the natural units with $\hbar=c=1$. From the following result
\begin{subequations}
\begin{eqnarray}
  ~[Y_{M_1 M_2\cdots M_{q+1}}]~
     &=& 
       [M^{(p+2)/2}]=(p+2)/2, \\
  ~[{\hat{Y}}^{(n)}_{\mu_1 \mu_2\cdots \mu_{q+1}}]~~~&=&
   [{\hat{Y}}^{(n)}_{\mu_1 \mu_2\cdots \mu_{q}z}]= (p+1)/2, \\
  ~[{\hat{X}}^{(n)}_{\mu_1\cdots \mu_q}]~~~~~ &=&
   [\hat{X}^{(n)}_{\mu_1\cdots\mu_{q-1}}]=(p-1)/2,\\
   ~ [U^{(n)}_1]=[U^{(n)}_2]&=&1/2,
\end{eqnarray}
\end{subequations}
we have
\begin{eqnarray}
  [I^{(3)}_{nn'}] =2,
  ~~[I^{(4)}_{nn'}]=1.\label{dimension1}
\end{eqnarray}

Further from the action {(\ref{actioneffective})}, the equations of motion for the effective fields can be obtained as
\begin{eqnarray}\label{effective2}
 \!\!\!\!\!\!\!\frac{1}{\sqrt{-\hat{g}}}\partial_{\mu_1} \left(\sqrt{-\hat{g}}\hat{Y}_{(n')}^{\mu_1{\mu_2...\mu_{q+1}}}\right)
 -\sum_n \Big( I^{(3)}_{nn'}\hat{X}_{(n)}^{{ \mu_2...\mu_{q+1}}}
             + I^{(4)}_{nn'}\hat{Y}_{(n)}^{{ \mu_2...\mu_{q+1}}}\Big)
=0,
\end{eqnarray}
and
\begin{eqnarray}\label{effective1}
 {\sum_{n'}\;I^{(2)}_{nn'}}\; \partial_{\mu_1} \left(\sqrt{-\hat{g}}\;\hat{Y}_{(n)}^{\mu_1\mu_2...\mu_{q}}\right)
  +\sum_{n'} {I^{(4)}_{nn'}}
         \partial_{\mu_1}
         \left(\sqrt{-\hat{g}}\;\hat{X}_{(n')}^{\mu_1\mu_2...\mu_{q}}\right)=0.
\end{eqnarray}

On the other hand, substituting the KK decomposition (\ref{KKdecomposition}) into Eqs. (\ref{equ1}) and (\ref{equ2}), we get
\begin{eqnarray}\label{effective01}
 \frac{1}{\sqrt{-\hat{g}}}
    \partial_{\mu_1}
    \left(\sqrt{-\hat{g}}\;\hat{Y}_{(n)}^{\mu_1 \mu_2...\mu_{q+1}}\right)
 +  \lambda_{1} \hat{X}_{(n)}^{\mu_2...\mu_{q+1}}
 + \lambda_{2} \hat{Y}_{(n)}^{ \mu_2...\mu_{q+1}}
 =0,
\end{eqnarray}
and
\begin{eqnarray}\label{effective02}
  \partial_{\mu_1} \left(\sqrt{-\hat{g}}\;\hat{Y}_{(n)}^{\mu_1\mu_2...\mu_{q}}\right)
  +\lambda_{3} \partial_{\mu_1}
        \left(\sqrt{-\hat{g}}\;\hat{X}_{(n)}^{\mu_1\mu_2...\mu_{q}}\right)
  =0,
\end{eqnarray}
where
\begin{subequations}\label{lambda}
\begin{eqnarray}
  \lambda_{1}&=&\frac{\text{e}^{-(a_1+p-2q)A}}
  {{(q+1)}\;U^{(n)}_1}\;
  \partial_z\bigg(\text{e}^{(p-2q)A}\;\partial_z\big(U^{(n)}_1\;\text{e}^{a_1A}\big)\bigg),
  \label{lambda_1}\\
  \lambda_{2}&=&\frac{q\;\text{e}^{-(a_1+p-2q)A}}{{(q+1)}\;U^{(n)}_1}\;
  \partial_z\bigg(U^{(n)}_2\;\text{e}^{(a_2+p-2q)A}\bigg), \label{lambda_2}  \\
  \lambda_{3}&=& \frac{\partial_z(U^{(n)}_1\;\text{e}^{a_1A})}
                      {q\;U^{(n)}_2\;\text{e}^{a_2A}}.  \label{lambda_3}
\end{eqnarray}
\end{subequations}
It is clear that Eqs. (\ref{effective2}), (\ref{effective1}) and (\ref{effective01}), (\ref{effective02}) must be consistent with each other, which results in that
\begin{eqnarray}
  I^{(3)}_{nn'}&=&\frac{m_{n}^2}{{q+1}}\;I^{(1)}_{nn'}
  =\frac{m_{n}^2}{{q+1}} {\delta_{nn'} }, \label{i311}\\
  I^{(4)}_{nn'}&=& \bar{m}_n \;\delta_{nn'},  \label{i411}
\end{eqnarray}
and
\begin{subequations}\label{lambda123}
\begin{eqnarray}
  \lambda_{1}&=&-\frac{m_{n}^2}{{q+1}},
     \label{uu3}\\
  \lambda_{2}&=&-\bar{m}_n  ,\label{uu2}\\
  \lambda_{3}&=& \frac{\bar{m}_n}{{q+1\;}}, \label{uu1}
\end{eqnarray}
\end{subequations}
where $[m_{n}]=[\bar{m}_n]=1$.

The above three equations are interesting. First, Eq. (\ref{uu3}) is in fact a Schr\"{o}dinger-like equation of $U_1^{(n)}$. With Eq. (\ref{uu1}) the expressions (\ref{condition2}) and (\ref{condition4}) are found to have the following relationship:
\begin{eqnarray}
  I^{(2)}_{nn'}&=&\frac{{(q+1)^2}}{\bar{m}_n^2}
  \frac{m_{n}^2}{{q+1}}
  \;\delta_{nn'} ={(q+1)}\;\delta_{nn'} ,\label{i21}\\
  I^{(4)}_{nn'}&=&\frac{m_{n}^2}{{\bar{m}_n}}
  \;\delta_{nn'} . \label{i41}
\end{eqnarray}
The consistency of $I^{(4)}_{nn'}$ in Eqs. (\ref{i411}) and (\ref{i41}) results in
\begin{equation}\label{rkm}
 \bar{m}_n^2={m_n^2}.
\end{equation}
{ Here we can choose $\bar{m}_n={+m_n}$ for the $q-$form field. But for the $(p-q)-$form field, which is dual to the $q-$form field through the Hodge duality, it should be the negative $-m_n$, because this is the requirement of {a new duality} on the brane. We will discuss this in Sec. \ref{dualitymassive}.} Then Eqs. (\ref{lambda_1})$\sim$(\ref{lambda_3}) and (\ref{uu3})$\sim$(\ref{uu1}) are equivalent to the following coupled equations of $U^{(n)}_1$ and $U^{(n)}_2$:
\begin{eqnarray}
  \partial_z U_2^{(n)}(z) + \frac{p-2q}{2}A'(z)\;U_2^{(n)}(z)
     &=& -\frac{q+1}{q} m_n  U_1^{(n)}(z),\label{sch11}\\
  \partial_z U_1^{(n)}(z)-\frac{p-2q}{2 }A'(z)\;U_1^{(n)}(z)
     &=& +\frac{q}{q+1} m_n  U_2^{(n)}(z),\label{sch22}
\end{eqnarray}
which can also be written as a Schr\"{o}dinger-like equation for each mode,
\begin{subequations}\label{sch}
\begin{eqnarray}
  \big[-\partial_z^2+V_{q,1}(z)\big]U^{(n)}_1(z) &=&m_{n}^2\;U^{(n)}_1(z),\label{sch1}\\
  \big[-\partial_z^2+V_{q,2}(z)\big]U^{(n)}_2(z) &=&m_{n}^2\;U^{(n)}_2(z),\label{sch2}
\end{eqnarray}
\end{subequations}
where the effective potentials are given by
\begin{eqnarray}
V_{q,1}(z)&=&\frac{(p-2q)^2}{4}A'^2(z) +\frac{p-2q}{2}A''(z),\label{veff1}\\
V_{q,2}(z)&=&\frac{(p-2q)^2}{4}A'^2(z) -\frac{p-2q}{2}A''(z).\label{veff2}
\end{eqnarray}
It is worth noting that the above two equations (\ref{sch1}) and (\ref{sch2}) can be rewritten alternatively as
\begin{eqnarray}
  \mathcal{Q}\mathcal{Q}^{\dag} \;U^{(n)}_1(z)  &=& m_{n}^2 U^{(n)}_1(z),\label{sch111}\\
  \mathcal{Q}^{\dag} \mathcal{Q} \;U^{(n)}_2(z) &=&m_{n}^2U^{(n)}_2(z),\label{sch2222}
\end{eqnarray}
with the operator $\mathcal{Q}$ given by $\mathcal{Q}=\partial_z +\frac{p-2q}{2}A'(z)$.
So we have the following conclusions: (1) there is no eigenstate with negative eigenvalue; namely, we always have $m_{n}^2\ge 0$. (2) There is only one zero mode with $m_0=0$, $U^{(0)}_1$ or $U^{(0)}_2$, that can survive with the boundary condition $U^{(0)}_{1,2}(|z|\rightarrow\infty) \rightarrow 0$. (3) The two base functions $U^{(n)}_1$ and $U^{(n)}_2$ share the same mass spectrum except for $m_0=0$.

Now it is {clear} that by solving the two Schr\"{o}dinger-like equations (\ref{sch1}) and (\ref{sch2}) with the orthonormality conditions (\ref{condition1}) and (\ref{condition2}), we can find the mass spectrum of the KK modes that can be localized on the brane.

We usually classify the KK modes into massless and massive ones, as the former are regarded as the field that has been on the brane, and the latter are carrying the information of extra dimensions, which can be distinguished from the ones that have been on the brane. For a realistic brane world, these two types {of} KK modes are expected to be localized on the brane. For the massless mode, its analytical wave function can easily be obtained, so we can check whether it can be localized on the brane through the normalization condition. While for the massive KK modes, we usually have to use a numerical method to solve them from the Schr\"{o}dinger-like equations. There may also exist bound massive KK modes. In the following we will discuss the massless and massive ones separately.

\subsection{Massless KK modes}\label{massless}

For the massless KK modes $U^{(0)}_{1,2}$ with $m_{0}=0$,  the solutions can be obtained from Eqs. (\ref{sch1}) and (\ref{sch2}) or Eqs. (\ref{sch11}) and (\ref{sch22}):
  \begin{eqnarray}
  U^{(0)}_1(z)&=& N_1\;\text{e}^{+(p-2q)A/2},\label{u1}\\
  U^{(0)}_2(z)&=& N_2\;\text{e}^{-(p-2q)A/2},\label{u2}
  \end{eqnarray}
where $N_1$ and $N_2$ are the normalization constants. Their effective action reads
\begin{eqnarray}
  {S_{q,0}} &=&\int d ^{p+1}x\sqrt{-\hat{g}}\bigg(I^{(1)}_{00}\;\hat{Y}_{(0)}^{\mu_1\cdots \mu_{q+1}}\hat{Y}^{(0)}_{\mu_1\cdots \mu_{q+1}}+I^{(2)}_{00}\;\hat{Y}_{(0)}^{\mu_1\cdots \mu_{q}}\hat{Y}^{(0)}_{\mu_1\cdots \mu_{q}}\bigg),
\end{eqnarray}
where
\begin{eqnarray}\label{u10}
  { I^{(1)}_{q,00}} &=&N_1^2 \int dz \;\text{e}^{(p-2q)A},\\
  { I^{(2)}_{q,00}}&=&N_2^2 \int dz \;\text{e}^{-(p-2q)A}.
\end{eqnarray}
It can be seen that { $I^{(1)}_{q,00}$ and $I^{(2)}_{q,00}$} cannot be finite at the same time for a brane with infinite extra dimension.

Then with (\ref{u1}) and (\ref{u2}), for the bulk $q-$form and its dual $(p-q)-$form fields (the duality is built through the Hodge duality in the bulk), their KK decompositions now are:
\begin{subequations}\label{decompositions2}
\begin{eqnarray}\label{zerokk1}
  Y_{\mu_1\cdots \mu_{q+1}}(x^{\mu},z)
     &=&\hat{Y}^{(0)}_{\mu_1\cdots \mu_{q+1}}(x^{\mu})\;N_1,\\
  Y_{\mu_1\cdots \mu_{q}z}(x^{\mu},z)
     &=&{\hat{Y}^{(0)}_{\mu_1\cdots \mu_q}(x^{\mu})}\;\frac{q}{q+1}N_2\;\text{e}^{(2q-p)A(z)},\label{zerokk2}\\
  \sqrt{-g}\;\widetilde{Y}^{\mu_1\cdots \mu_{p-q+1}}(x^{\mu},z)
     &=&\sqrt{-\hat{g}}\; \widetilde{\hat{Y}}_{(0)}^{\mu_1\cdots \mu_{p-q+1}} (x^{\mu})
        \;\widetilde{N}_1\;\text{e}^{(2q-p)A(z)},\label{zerokk3}\\
  \sqrt{-g}\;\widetilde{Y}^{\mu_1\cdots \mu_{p-q}z} (x^{\mu},z)
     &=&\sqrt{-\hat{g}}\;\widetilde{\hat{Y}}_{(0)}^{\mu_1\cdots \mu_{p-q}} (x^{\mu})
         \;\frac{p-q}{p-q+1}\widetilde{N}_2\label{zerokk4},
\end{eqnarray}
\end{subequations}
where we have supposed that there is only the zero mode that is localized on the brane. Then substituting the above decompositions (\ref{decompositions2}) into the below bulk Hodge duality \cite{Duff2000se}
{\begin{eqnarray}
  \sqrt{-g}\;{\widetilde{Y}^{M_1\cdots M_{p-q+1}}}&=&{\frac{1}{(q+1)!}}\;\varepsilon ^{M_1\cdots M_{p-q+1}N_1\cdots N_{q+1}}{Y_{N_1\cdots N_{q+1}}},\label{HodgeDualbulkt}
\end{eqnarray}
which can also be written as
\begin{subequations}\label{HodgeDualbulk}
\begin{eqnarray}
  \sqrt{-g}\;{\widetilde{Y}^{\mu_1\cdots \mu_{p-q}z}}&=&
  {\frac{1}{(q+1)!}}\;
  \varepsilon ^{\mu_1\cdots \mu_{p-q}z\nu_1\cdots \nu_{q+1}}{Y_{\nu_1\cdots \nu_{q+1}}},\label{HodgeDualbulk1}\\
  \sqrt{-g}\;{\widetilde{Y}^{\mu_1\cdots \mu_{p-q+1}}}&=&{\frac{1}{q!}}\;
  \varepsilon ^{\mu_1\cdots \mu_{p-q+1}\nu_1\cdots \nu_{q}z}\;{Y_{\nu_1\cdots \nu_{q}z}},\label{HodgeDualbulk2}
\end{eqnarray}
\end{subequations}}
the Hodge duality on the brane is naturally satisfied:
\begin{subequations}\label{HodgeDualonBrane}
\begin{eqnarray}
\sqrt{-\hat{g}}\;\widetilde{\hat{Y}}_{(0)}^{\mu_1\cdots \mu_{p-q}}(x^{\mu})
 &=&{\frac{1}{(q+1)!}}\;\varepsilon^{\mu_1\cdots \mu_{p-q}\nu_1\cdots \nu_{q+1}}\;{\hat{Y}^{(0)}_{\nu_1\cdots \nu_{q+1}}}(x^{\mu}),\label{HodgeDualBrane1}\\
 \sqrt{-\hat{g}}\;{\widetilde{\hat{Y}}_{(0)}^{\mu_1\cdots \mu_{p-q+1}}}(x^{\mu})
 &=&{\frac{1}{q!}}\;
 \varepsilon^{\mu_1\cdots \mu_{p-q+1}\nu_1\cdots \nu_{q}}
 \;\hat{Y}^{(0)}_{\nu_1\cdots \nu_q}(x^{\mu}), \label{HodgeDualBrane2}
\end{eqnarray}
\end{subequations}
where we have assumed that
\begin{eqnarray}\label{norcond}
  N_1=\frac{p-q}{p-q+1}\widetilde{N}_2,
  ~~~N_2=\frac{q+1}{q}\widetilde{N}_1.
\end{eqnarray}
{From} (\ref{HodgeDualBrane1}) we see that there is a duality between a massless $q-$form field and a massless $(p-q-1)-$form one on the brane:
{
\begin{eqnarray}\label{masslessdual1}
{S_{q,0}}
   &=&{-\frac{1}{2(q+1)!}}\;
     \int d^{p+1}x \sqrt{-\hat{g}}\;\hat{Y}_{(0)}^{\mu_1\cdots \mu_{q+1}}
           \hat{Y}^{(0)}_{\mu_1\cdots \mu_{q+1}},
           \nonumber\\
    &=&\nonumber\\
{\widetilde{S}_{p-q-1,0}}
   &=&{-\frac{1}{2(p-q)!}}
     \int d^{p+1}x \sqrt{-\hat{g}}
     \;\widetilde{\hat{Y}}_{(0)}^{\mu_1\cdots \mu_{p-q}}
           \widetilde{\hat{Y}}^{(0)}_{\mu_1\cdots \mu_{p-q}},
\end{eqnarray}
From (\ref{HodgeDualBrane2}), a massless $(q-1)-$form field is dual to a massless $(p-q)-$form one:
\begin{eqnarray}\label{masslessdual2}
{S_{q-1,0}}
   &=&{-\frac{1}{2q!}}\;
     \int d^{p+1}x \sqrt{-\hat{g}}\;\hat{Y}_{(0)}^{\mu_1\cdots \mu_{q}}
           \hat{Y}^{(0)}_{\mu_1\cdots \mu_{q}},
           \nonumber\\
    &=&\nonumber\\
{\widetilde{S}_{p-q,0}}
   &=&{-\frac{1}{2(p-q+1)!}}
     \int d^{p+1}x \sqrt{-\hat{g}}
     \;\widetilde{\hat{Y}}_{(0)}^{\mu_1\cdots \mu_{p-q+1}}
           \widetilde{\hat{Y}}^{(0)}_{\mu_1\cdots \mu_{p-q+1}}.
\end{eqnarray}}

From the discussion about the localization of the zero mode for a $q-$form field, we see that there is also a $(p-q)-$form or $(p-q-1)-$form zero mode for a bulk $(p-q)-$form field. It is interesting to note that for a $q-$form field and its dual $(p-q)-$form field, with Eq. (\ref{norcond}), there are some relationships between the normalization constants:
\begin{eqnarray}
{ \widetilde{I}^{(1)}_{p-q,00}} &=&
 \widetilde{N}_1^2\int dz\;\text{e}^{(p-2(p-q))A}={ \frac{I^{(2)}_{q,00}}{q+1}},\\
{ \widetilde{I}^{(2)}_{p-q,00}} &=& \frac{(p-q)^2}{p-q+1}\;
\widetilde{N}_2^2\int dz\;\text{e}^{(2(p-q)-p)A}={(p-q+1)\;I^{(1)}_{q,00}},
\end{eqnarray}
where {$\widetilde{I}^{(1)}_{p-q,00}$ and $\widetilde{I}^{(2)}_{p-q,00}$} are the normalization constants appearing in the effective action of the $(p-q)-$form field. It is clear that if there is a localized $q-$form ( or $(q-1)-$form ) zero mode for a bulk $q-$form field, there must be a localized $(p-q-1)-$form ( or $(p-q)-$form ) zero mode for its dual field. And this just satisfies the requirement of the Hodge duality on the brane (\ref{HodgeDualonBrane}).

We have known that on some $3-$brane models there is a localized $1-$form zero mode \cite{Fu:2013ita} from a bulk $1-$form field, but now it is seen that the localized $1-$form also may be from a bulk $2-$form one. In fact, for any massless effective $q-$form field dual to itself on the brane, we {cannot be sure} it is from a bulk $q-$form or $(p-q)-$form field.

\subsection{ Bound massive KK modes}\label{massive}
Furthermore, for some brane backgrounds there may be bound massive KK modes except for the localized zero mode. In this case we are wondering if we substitute the general KK decomposition into the Hodge duality in the bulk and keep the Hodge duality on the brane valid, what will happen for the bound massive KK modes for the $q-$form and its dual fields. Let us consider this issue in the following discussion.

For the bound massive KK modes, as the two Schr\"{o}dinger-like equations (\ref{sch1}) and (\ref{sch2}) are not independent of each other, we can get the mass spectra from any one of them with the corresponding orthonormality condition (\ref{condition1}) or (\ref{condition2}). The effective action for these $n-$level bound KK modes {of the $q$-form field} can be written as
\begin{eqnarray}\label{actionmass}
{S_{q,n}}
   &=&{-\frac{1}{2(q+1)!}}
     \int d^{p+1}x \sqrt{-\hat{g}}
     \;\hat{Y}_{(n)}^{\mu_1\cdots \mu_{q+1}}
           \hat{Y}^{(n)}_{\mu_1\cdots \mu_{q+1}}\nonumber\\
    && {-\frac{1}{2q!}}\;
    \int d^{p+1}x \sqrt{-\hat{g}}\;
    {\bigg(\hat{Y}_{(n)}^{\mu_1\cdots \mu_{q}}
           + \frac{m_n}{q+1}\hat{X}_{(n)}^{\mu_1\cdots \mu_{q}}\bigg)^2}
           .
\end{eqnarray}
Different with the zero modes, which have two types ($q-$form and $(q-1)-$form), the bound massive KK modes are all $q-$form fields, and each $n-$level bound massive $q-$form KK mode couples with the $n-$level massless $(q-1)-$form mode.

{ This coupling is important. Because of it the effective action (\ref{actionmass}) is in fact gauge {invariant}  under the {following} gauge transformation:
\begin{eqnarray}
   \hat{X}_{\mu_1\cdots \mu_q}^{(n)} &\rightarrow& \hat{X}_{\mu_1\cdots \mu_q}^{(n)}+\partial_{\mu_1}\hat{\Lambda}_{\mu_2\cdots \mu_{q}},\\
   {\hat{X}_{\mu_2\cdots \mu_{q}}^{(n)} } &\rightarrow&
      {\hat{X}_{\mu_2\cdots \mu_{q}}^{(n)}}
      - \frac{m_n}{q+1}\hat{\Lambda}_{\mu_2\cdots \mu_{q}},
\end{eqnarray}
where $\hat{\Lambda}_{\mu_2\cdots \mu_{q}}$ is an antisymmetric tensor field. For {a} $1-$form field, the effective action is just the St$\ddot{u}$ckelberg one. Then we can fix the gauge by choosing
\begin{eqnarray}\label{gauge2}
  \partial^{\mu_1}\hat{X}_{\mu_1\cdots \mu_q}^{(n)}=0,
\end{eqnarray}
{under which we will simplify the action (\ref{actionmass}) and obtain the physical {masses} of the massive bound KK modes.}

{With the definition  }
\begin{eqnarray}
  \hat{Y}_{\mu_1\cdots \mu_{q+1}}^{(n)}=\frac{1}{q+1}
  \bigg(\partial_{\mu_1}\hat{X}_{\mu_2\mu_3\cdots\mu_{q+1}}^{(n)}
  +\partial_{\mu_2}\hat{X}_{\mu_3\cdots\mu_{q+1}\mu_1}^{(n)}
  +\partial_{\mu_3}\hat{X}_{\mu_4\cdots\mu_{q+1}\mu_1\mu_2}^{(n)}
  +\cdots\bigg),
\end{eqnarray}
{we have }
\begin{eqnarray}
  \hat{Y}_{\mu_1\cdots \mu_{q+1}}^{(n)}\hat{Y}^{\mu_1\cdots \mu_{q+1}}_{(n)}&=&
  \frac{1}{q+1}\partial_{\mu_1}\hat{X}_{\mu_2\mu_3\cdots\mu_{q+1}}^{(n)}
  \partial^{\mu_1}\hat{X}^{\mu_2\mu_3\cdots\mu_{q+1}}_{(n)}\nonumber\\
  &&+\; \frac{1}{(q+1)^2}
  \bigg(\partial_{\mu_1}\hat{X}_{\mu_2\mu_3\cdots\mu_{q+1}}^{(n)}
  \partial^{\mu_2}\hat{X}^{\mu_3\cdots\mu_{q+1}\mu_1}_{(n)}+\cdots\bigg).
\end{eqnarray}
{ By using the gauge condition (\ref{gauge2}), the above expression can be written as}
\begin{eqnarray}
  \hat{Y}_{\mu_1\cdots \mu_{q+1}}^{(n)}\hat{Y}^{\mu_1\cdots \mu_{q+1}}_{(n)}={-}\frac{1}{q+1}\hat{X}_{\mu_2\mu_3\cdots\mu_{q+1}}^{(n)}
  \partial_{\mu_1}\partial^{\mu_1}\hat{X}^{\mu_2\mu_3\cdots\mu_{q+1}}_{(n)}
  +\text{total derivative terms }.
\end{eqnarray}
While the term $\hat{Y}_{\mu_1\cdots \mu_{q}}^{(n)}\hat{X}^{\mu_1\cdots \mu_{q}}_{(n)}$ is in fact a {total derivative} one. Therefore, under the gauge condition (\ref{gauge2}), the effective action (\ref{actionmass}) for the $n-$level bound KK mode of the $q-$form field {turns out} to be
\begin{eqnarray}
  S_{q,n}^{\text{gauged}}
   &=&-{\frac{1}{2(q+1)!}}
     \int d^{p+1}x \sqrt{-\hat{g}}
    \bigg[-\frac{1}{q+1}
         \Big(\hat{X}_{\mu_1\cdots\mu_{q}}^{(n)}
          \hat{\Box}~ \hat{X}^{\mu_1\cdots\mu_{q}}_{(n)}
           -m_{n}^2
          \hat{X}_{\mu_1\cdots\mu_{q}}^{(n)}
          \hat{X}^{\mu_1\cdots \mu_{q}}_{(n)}\Big)\nonumber\\
    &&~~~~~~~~~~~~~~~~~~~~~~~~~~~~~~~~~
    +{(q+1)\;}\hat{Y}_{\mu_1\cdots \mu_q}^{(n)}\hat{Y}^{\mu_1\cdots \mu_q}_{(n)} \bigg],
\end{eqnarray}
where $\hat{\Box}=\eta^{\mu\nu}\partial_{\mu}\partial_{\nu}$. Note that here we have considered the case of a flat brane, i.e., $\hat{g}_{\mu\nu}=\eta_{\mu\nu}$. Now it is clear that the $n-$level massive KK mode $\hat{X}^{\mu_1\cdots\mu_{q}}_{(n)}$ (a massive $q-$form {field} on the brane) decouples {from} the $n-$level massless KK mode $\hat{X}^{\mu_1\cdots\mu_{q-1}}_{(n)}$ (a massless $(q-1)-$form {field} on the brane). There is still a gauge freedom for the $n-$level massless KK mode.

The equation of motion for the $n-$level massive KK mode {}is
\begin{eqnarray}\label{equationm}
  \left(\hat{\Box}-m_{n}^2\right)\hat{X}^{\mu_1\cdots\mu_{q}}_{(n)}=0.
\end{eqnarray}
Neglecting the tensor structure, the propagator for the mode can be calculated as
\begin{eqnarray}
  G(x,y)=\int \frac{d^{p+1}k}{(2\pi)^{p+1}}\;\frac{i~ \text{e}^{-ik(x-y)}}{k^2+m^2_n},
\end{eqnarray}
which is the Green's function for (\ref{equationm}) with $k$ the momentum of the mode. {Therefore,} the physical mass for the KK {mode} is
\begin{equation}
  m_{\text{phy}}^2=m_{n}^2.
\end{equation}
}

{On the other hand, we find that} the effective potentials of the $q-$form and its dual $(p-q)-$form fields have the following relationships:
\begin{equation}\label{relationshipvqvdq}
  \widetilde{V}_{p-q,1}(z)=V_{q,2}(z),~~~~\widetilde{V}_{p-q,2}(z)=V_{q,1}(z).
\end{equation}
Therefore, {for the $(p-q)-$form field there also exist bound massive $(p-q)-$form KK modes with the same mass spectra $m_{n}$ and the corresponding massless $(p-q-1)-$form modes}. So that we could say that the {$n-$level ($n>0$)} bound KK modes for the $q-$form and $(p-q)-$form field are dual to each other. Next, we will prove this duality by considering the Hodge duality in the bulk and derive the duality between them on the brane.

\subsection{{A new} duality on the brane}\label{dualitymassive}

From the relationships (\ref{relationshipvqvdq}), we have $\widetilde{U}_1^{(n)}=c_1 U_2^{(n)}$, $\widetilde{U}_2^{(n)}=c_2 U_1^{(n)}$, where $\widetilde{U}_1^{(n)}$ and  $\widetilde{U}_2^{(n)}$ are two functions {appearing} in the KK decompositions of the dual $(p-q)-$form {field}:
\begin{subequations}\label{KKdecompositionDualfield}
\begin{eqnarray}
  \widetilde{X}_{\mu_1 \cdots \mu_{p-q}(x_{\mu},z)} &=& \sum_{n}\widetilde{\hat{X}}_{\mu_1\cdots \mu_{p-q}}^{(n)}(x^{\mu})\;\widetilde{U}_1^{(n)}(z)\text{e}^{\widetilde{a}_1A(z)}, \label{kk3} \\
  \widetilde{X}_{\mu_1 \cdots \mu_{p-q-1}z}(x_{\mu},z) &=& \sum_{n}\widetilde{\hat{X}}_{\mu_1\cdots \mu_{p-q-1}}^{(n)}(x^{\mu})\;\widetilde{U}_2^{(n)}(z)\text{e}^{\widetilde{a}_2A(z)}.\label{kk4}
\end{eqnarray}
\end{subequations}
Here $\widetilde{a}_1=\widetilde{a}_2=(p-2q)/2$.
The effective action for the dual $(p-q)-$form field corresponding to (\ref{actioneffective}) reads
\begin{eqnarray}\label{actioneffectivedual}
  \widetilde{S}_{p-q} 
   &=&{-\frac{1}{2(p-q+1)!}}
     \sum_n\sum_{n'}\bigg[ \widetilde{I}^{(1)}_{nn'}\int d^{p+1}x \sqrt{-\hat{g}} \;
           \widetilde{\hat{Y}}_{(n)}^{\mu_1\cdots \mu_{p-q+1}}
           \widetilde{\hat{Y}}^{(n')}_{\mu_1\cdots \mu_{p-q+1}} \nonumber\\
   &&\quad\quad\quad+ \widetilde{I}^{(2)}_{nn'}\int d^{p+1}x \sqrt{-\hat{g}}\;
           \widetilde{\hat{Y}}_{(n)}^{\mu_1\cdots \mu_{p-q}}
           \widetilde{\hat{Y}}^{(n')}_{\mu_1\cdots \mu_{p-q}}\nonumber\\
   &&\quad\quad\quad+ \widetilde{I}^{(3)}_{nn'}\int d^{p+1}x \sqrt{-\hat{g}}\;
           \widetilde{\hat{X}}_{(n)}^{\mu_1\cdots \mu_{p-q}}
           \widetilde{\hat{X}}^{(n')}_{\mu_1\cdots \mu_{p-q}}\nonumber\\
   &&\quad\quad\quad+  2\widetilde{I}^{(4)}_{nn'}\int d ^{p+1}x\sqrt{-\hat{g}}\;
           \widetilde{\hat{Y}}_{(n)}^{\mu_1\cdots \mu_{p-q}}
           \widetilde{\hat{X}}^{(n')}_{\mu_1\cdots \mu_{p-q}}\bigg].
      \label{Sp-q}
\end{eqnarray}
 Then with the relations $\widetilde{I}^{(1)}_{nn'} \equiv \int dz\;\widetilde{U}_1^{(n)}\widetilde{U}_1^{(n')}
    =\delta_{nn'}$, $\widetilde{U}_1^{(n)}=c_1 U_2^{(n)}$, and (\ref{condition2}), we have $c_1=q/(q+1)$. Similarly, we get $c_2=(p-q+1)/(p-q)$. Thus, we obtain the following relations:
\begin{subequations}\label{relationsU}
 \begin{eqnarray}
  \widetilde{U}_1^{(n)} &=& \frac{q}{q+1} U_2^{(n)}, \label{relationsUa}\\
  \widetilde{U}_2^{(n)} &=& \frac{p-q+1}{p-q} U_1^{(n)}. \label{relationsUb}
\end{eqnarray}
\end{subequations}

{With the two above relations, and considering the two coupled equations of $U^{(n)}_1$ and $U^{(n)}_2$ (\ref{sch11}) and (\ref{sch22}), we can find that
\begin{eqnarray}
  \partial_z \widetilde{U}_2^{(n)}(z) - \frac{p-2q}{2}A'(z)\;\widetilde{U}_2^{(n)}(z)
     &=& \frac{p-q+1}{p-q} m_n  \widetilde{U}_1^{(n)}(z),\label{sch11dual}\\
  \partial_z \widetilde{U}_1^{(n)}(z)+\frac{p-2q}{2 }A'(z)\;\widetilde{U}_1^{(n)}(z)
     &=& -\frac{p-q}{p-q+1} m_n  \widetilde{U}_2^{(n)}(z).\label{sch22dual}
\end{eqnarray}
While the Eq. (\ref{sch22dual}) is just equal to
\begin{eqnarray}\label{tildelamda}
\frac{\partial_z(\widetilde{U}^{(n)}_1\;\text{e}^{\widetilde{a}_1A})}
{(p-q)\;\widetilde{U}_2^{(n)}(z)
 \;\text{e}^{\widetilde{a}_2A}} &=&
 -\frac{m_n}{p-q+1}=\frac{\widetilde{I}^{(4)}_{nn'}}{p-q+1},
\end{eqnarray}
which is similar to the one for the $q-$form field (\ref{lambda_3}). This means that for the $(p-q)-$form field, we have
\begin{equation}
  \widetilde{I}^{(4)}_{nn'}=-m_n\delta_{nn'},
\end{equation}
so that for the massive bound KK modes of the $(p-q)-$form field, the effective action is
\begin{eqnarray}\label{massiveactiondual}
  {\widetilde{S}_{p-q,n}}
   &=&{-\frac{1}{2(p-q+1)!}}
     \int d^{p+1}x \sqrt{-\hat{g}}
     \;\widetilde{\hat{Y}}_{(n)}^{\mu_1\cdots \mu_{p-q+1}}
           \widetilde{\hat{Y}}^{(n)}_{\mu_1\cdots \mu_{p-q+1}}\nonumber\\
          &&
     {-\frac{1}{2(p-q)!}}
     \int d^{p+1}x \sqrt{-\hat{g}}\left(\widetilde{\hat{Y}}_{(n)}^{\mu_1\cdots \mu_{p-q}}{-}\frac{m_n}{p-q+1}\widetilde{\hat{X}}_{(n)}^{\mu_1\cdots \mu_{p-q}}\right)^2.
\end{eqnarray}
According to the calculations about the propagators for the $q-$form field, we can see that the physical masses for massive KK modes of the $(p-q)-$form field are also $m_n$.}

{{With the KK decompositions of the bulk $q-$form and its dual fields  (\ref{KKdecomposition}) and (\ref{KKdecompositionDualfield}), and {considering the relations (\ref{lambda_3}) and (\ref{tildelamda})}, the KK decompositions of the field strengths} can be written as}
\begin{subequations}\label{KKdecompositionFieldStrength}
\begin{eqnarray}
Y_{\nu_1\cdots\nu_{q+1}} (x^{\mu},z)&=&
     N_1\hat{Y}^{(0)}_{\nu_1\cdots\nu_{q+1}}(x^{\mu})
     +\text{e}^{a_1A}\;
     \sum_{n\ge1}\hat{Y}_{\nu_1\cdots\nu_{q+1}}^{(n)} (x^{\mu}) U_1^{(n)}(z), \label{KKFieldStrength1}\\
Y_{\nu_1\cdots\nu_{q}z}(x^{\mu},z) &=&
     {\hat{Y}^{(0)}_{\nu_1\cdots\nu_q}(x^{\nu})}
     \;\widetilde{N}_1\;\text{e}^{(2q-p)A(z)}\nonumber\\
    &+&\frac{q}{q+1} \text{e}^{a_2A}\;\sum_{n\ge1} \bigg(\hat{Y}_{\nu_1\cdots\nu_{q}}^{(n)}(x^{\mu})
      +\frac{m_n}{q+1}\hat{X}_{\nu_1\cdots\nu_{q}}^{(n)}(x^{\mu})\bigg)U^{(n)}_2(z),\label{KKFieldStrength2}
\end{eqnarray}
\end{subequations}
\begin{subequations}\label{{KKdecompositionDualFieldStrength}}
\begin{eqnarray}
\sqrt{-g}\;\widetilde{Y}^{\mu_1\cdots\mu_{p-q+1}} (x^{\mu},z) &=&
   \sqrt{-\hat{g}}\;\bigg[\widetilde{\hat{Y}}_{(0)}^{\mu_1\cdots\mu_{p-q+1}}(x^{\mu})\;
    \widetilde{N}_1\text{e}^{(2q-p)A}
    \nonumber\\
    &+&\text{e}^{(\widetilde{a}_1+2q-p)A}
    \;\sum_{n\ge1}\widetilde{\hat{Y}}^{\mu_1\cdots\mu_{p-q+1}}_{(n)}(x^{\mu})
    \;\widetilde{U}_1^{(n)}(z) \bigg], \label{KKDualFieldStrength1} \\
\sqrt{-g}\;\widetilde{Y}^{\mu_1\cdots\mu_{p-q}z} (x^{\mu},z)&=&
   \sqrt{-\hat{g}}\;\bigg[N_1\;{\widetilde{\hat{Y}}_{(0)}^{\mu_1\cdots \mu_{p-q}}}(x^{\mu}) +\frac{p-q}{p-q+1}\text{e}^{(\widetilde{a}_2+2q-p)A}
   \nonumber\\
  &\times&\sum_{n\ge1}\Big(\widetilde{\hat{Y}}^{\mu_1\cdots\mu_{p-q}}_{(n)}(x^{\mu})
  {-}\frac{m_n}{p-q+1}\widetilde{\hat{X}}^{\mu_1\cdots\mu_{p-q}}_{(n)}(x^{\mu})\Big)\widetilde{U}^{(n)}_2(z)\bigg].
   \label{KKDualFieldStrength2}
\end{eqnarray}
\end{subequations}

{Substituting the above field decompositions (\ref{KKDualFieldStrength2}) and (\ref{KKFieldStrength1}) into the Hodge duality (\ref{HodgeDualbulk}) and considering the Hodge duality of the zero modes on the brane (\ref{HodgeDualBrane1}) as well as the relations (\ref{relationsU}), we obtain the following dual relation between the $n-$level KK modes:
\begin{eqnarray}
 \sqrt{-\hat{g}}\;
 \left(\widetilde{\hat{Y}}_{(n)}^{\mu_1\cdots \mu_{p-q}}{-}\frac{m_n}{p-q+1}\widetilde{\hat{X}}_{(n)}^{\mu_1\cdots \mu_{p-q}}\right)
 ={\frac{1}{(q+1)!}}\;\varepsilon^{\mu_1\cdots \mu_{p-q}\nu_1\cdots \nu_{q+1}}\;{\hat{Y}^{(n)}_{\nu_1\cdots \nu_{q+1}}}.\label{HodgeDualityOnBrane1}
\end{eqnarray}
Similarly, we can derive another dual relation:
\begin{eqnarray}
 \sqrt{-\hat{g}}\;{\widetilde{\hat{Y}}_{(n)}^{\mu_1\cdots \mu_{p-q+1}}}
 ={\frac{1}{q!}}\;
 \varepsilon^{\mu_1\cdots \mu_{p-q+1}\nu_1\cdots \nu_{q}}
 \Big(\hat{Y}^{(n)}_{\nu_1\cdots \nu_q}+\frac{m_n}{q+1}\hat{X}^{(n)}_{\nu_1\cdots \nu_q}\Big). \label{HodgeDualityOnBrane2}
\end{eqnarray}
With the above two relations it is easy to show that the two effective actions on the brane of the $n-$level KK modes of the bulk $q-$form and its dual $(p-q)-$form  fields are equal:
\begin{eqnarray}\label{smass}
  S_{q,n}&=& \widetilde{S}_{p-q,n}~.
\end{eqnarray}
This means that a bound $n-$level massive $q-$form KK mode coupling with an $n-$level massless $(q-1)-$form mode is dual to a bound $n-$level massive $(p-q)-$form one coupling with an $n-$level massless $(p-q-1)-$form mode. {The duality} relations are given by Eqs. (\ref{HodgeDualityOnBrane1}) and (\ref{HodgeDualityOnBrane2}). If we consider the gauge condition (\ref{gauge2}) for the $n-$level KK modes of the $q-$form field and the corresponding condition for the $n-$level KK modes of the dual $(p-q)-$form field, then we will find that a bound $n-$level $q-$form KK mode with mass $m_n$ and an independent $n-$level massless $(q-1)-$form mode are dual to a bound $n-$level massive $(p-q)-$form one with the same mass and an independent $n-$level massless $(p-q-1)-$form mode.}


\section{Conclusion and discussion}
\label{conclusion}

In this work, we investigated a new localization mechanism for a massless $q-$form field with a general KK decomposition without any gauge choice. It was found that for the KK modes of the $q-$form field, there are two Schr\"{o}dinger-like equations. By solving these two equations we can obtain the mass spectra of the KK modes and analyze their characters.

We found that for any $q-$form field there are two types of zero modes, a $q-$form mode and a $(q-1)-$form mode, which cannot be localized on the $p-$brane at the same time. We also found that an $n-$level massive bound KK mode couples with an $n-$level massless $(q-1)-$form mode, and both of them may localize on some $p-$branes. This suggests that if there {are} bound $n-$level massive vector KK modes for a bulk vector field, they must couple with $n-$level massless scalar {fields}. This is similar to the Higgs mechanism for a massless vector field obtaining mass; the difference is that the scalar field here is a part of the bulk vector field. By analyzing the gauge invariant effective action of the $n-$level KK modes and choosing a gauge condition, the $n-$level massive $q-$form KK mode decouples {from} the $n-$level massless $(q-1)-$form one.

Considering the general KK decomposition, we also found that the Hodge duality in the bulk naturally becomes two dualities on the brane. The first one is the Hodge duality between a $q-$form zero mode and a $(p-q-1)-$form one,  or between a $(q-1)-$form zero mode and a $(p-q)-$form one, which are indicated in (\ref{HodgeDualBrane1}) and (\ref{HodgeDualBrane2}).
The second duality is between two group KK modes: one is an $n-$level $q-$form KK mode with mass $m_n$ and an $n-$level massless $(q-1)-$form mode; another is an $n-$level $(p-q)-$form one with the same mass $m_n$ and an $n-$level massless $(p-q-1)-$form mode, {which are suggested in (\ref{HodgeDualityOnBrane1}) and (\ref{HodgeDualityOnBrane2}).}
The first kind of duality  is the usual Hodge duality between two massless {form} fields. But the second one is a new type of duality between two groups of form fields including massive and massless form fields. Note that the prerequisite for those dualities is that the corresponding KK modes should be localized on the brane. These dualities are listed in Table \ref{dualities}:
\begin{table}[h]
\begin{center}
\renewcommand\arraystretch{1.3}
\begin{tabular}
 {|l| c| c|}
  \hline
 \multicolumn{2}{|c|}{}&\textbf{Duality}\\
  \hline
\textbf{bulk} & \emph{massless}& $q-$form $\Leftrightarrow$ $(p-q)-$form\\
  \hline
\textbf{brane} & \emph{KK modes }&$q-$form $\Leftrightarrow$ $(p-q-1)-$form \\
& \emph{$n=0$}& or \\
& & $(q-1)-$form $\Leftrightarrow$ $(p-q)-$form\\
\cline{2-3}
& \emph{KK modes }& $q-$form(with mass $m_n$)+$(q-1)-$form(massless)\\
&\emph{$n\geqslant1$} & $\Updownarrow$ \\
& & $(p-q)-$form(with mass $m_n$)+$(p-q-1)-$form(massless)\\
  \hline
\end{tabular}
\end{center}
\caption{Dualities in the bulk and on the brane.}\label{dualities}
\end{table}

Because of the two dualities, there are some interesting phenomena. For example, since a $1-$form zero mode (a massless vector) on a $3-$brane is dual to itself according to the Hodge duality on the brane, we cannot judge whether fit is from a bulk $1-$form or a $2-$form field. In fact, { for any $q-$form zero mode dual to} itself on the $p-$brane, it may be from a bulk $q-$form or $(p-q)-$form one. {The same applies to the $n-$level KK modes. However, it does not matter since the effective field theories on the brane for the KK modes of the two dual bulk form fields are physically equivalent.}

\section{Acknowledgement}

We would like to thank the referee for his/her useful
comments and suggestions, which were very helpful in
improving this paper. This work was supported by the National Natural Science Foundation of China (Grants No. 11405121, No. 11375075, No. 11305119, and No. 11374237), and
the Fundamental Research Funds for the Central Universities (Grant No. lzujbky-2015-jl1). {C.-E. Fu was also supported by the scholarship granted by the Chinese Scholarship Council (CSC).}


\begin{thebibliography}{63}
\expandafter\ifx\csname natexlab\endcsname\relax\def\natexlab#1{#1}\fi
\expandafter\ifx\csname bibnamefont\endcsname\relax
  \def\bibnamefont#1{#1}\fi
\expandafter\ifx\csname bibfnamefont\endcsname\relax
  \def\bibfnamefont#1{#1}\fi
\expandafter\ifx\csname citenamefont\endcsname\relax
  \def\citenamefont#1{#1}\fi
\expandafter\ifx\csname url\endcsname\relax
  \def\url#1{\texttt{#1}}\fi
\expandafter\ifx\csname urlprefix\endcsname\relax\def\urlprefix{URL }\fi
\providecommand{\bibinfo}[2]{#2}
\providecommand{\eprint}[2][]{\url{#2}}

\bibitem[{\citenamefont{Antoniadis et~al.}(1998)\citenamefont{Antoniadis,
  Arkani-Hamed, Dimopoulos, and Dvali}}]{Antoniadis1998}
\bibinfo{author}{\bibfnamefont{I.}~\bibnamefont{Antoniadis}},
  \bibinfo{author}{\bibfnamefont{N.}~\bibnamefont{Arkani-Hamed}},
  \bibinfo{author}{\bibfnamefont{S.}~\bibnamefont{Dimopoulos}},
  \bibnamefont{and} \bibinfo{author}{\bibfnamefont{G.}~\bibnamefont{Dvali}},
  \bibinfo{journal}{Phys. Lett.} \textbf{\bibinfo{volume}{B 436}},
  \bibinfo{pages}{257} (\bibinfo{year}{1998}), \eprint{hep-ph/9804398}.

\bibitem[{\citenamefont{Randall and
  Sundrum}(1999{\natexlab{a}})}]{Randall1999a}
\bibinfo{author}{\bibfnamefont{L.}~\bibnamefont{Randall}} \bibnamefont{and}
  \bibinfo{author}{\bibfnamefont{R.}~\bibnamefont{Sundrum}},
  \bibinfo{journal}{Phys. Rev. Lett.} \textbf{\bibinfo{volume}{83}},
  \bibinfo{pages}{3370} (\bibinfo{year}{1999}{\natexlab{a}}),
  \eprint{hep-ph/9905221}.

\bibitem[{\citenamefont{Randall and
  Sundrum}(1999{\natexlab{b}})}]{Randall1999b}
\bibinfo{author}{\bibfnamefont{L.}~\bibnamefont{Randall}} \bibnamefont{and}
  \bibinfo{author}{\bibfnamefont{R.}~\bibnamefont{Sundrum}},
  \bibinfo{journal}{Phys. Rev. Lett.} \textbf{\bibinfo{volume}{83}},
  \bibinfo{pages}{4690} (\bibinfo{year}{1999}{\natexlab{b}}),
  \eprint{hep-th/9906064}.

\bibitem[{\citenamefont{Rubakov and Shaposhnikov}(1983)}]{Rubakov2}
\bibinfo{author}{\bibfnamefont{V.}~\bibnamefont{Rubakov}} \bibnamefont{and}
  \bibinfo{author}{\bibfnamefont{M.}~\bibnamefont{Shaposhnikov}},
  \bibinfo{journal}{Phys. Lett} \textbf{\bibinfo{volume}{B 125}},
  \bibinfo{pages}{139} (\bibinfo{year}{1983}).

\bibitem[{\citenamefont{Arkani-Hamed et~al.}(1998)\citenamefont{Arkani-Hamed,
  Dimopoulos, and Dvali}}]{ArkaniHamed1998rs}
\bibinfo{author}{\bibfnamefont{N.}~\bibnamefont{Arkani-Hamed}},
  \bibinfo{author}{\bibfnamefont{S.}~\bibnamefont{Dimopoulos}},
  \bibnamefont{and} \bibinfo{author}{\bibfnamefont{G.}~\bibnamefont{Dvali}},
  \bibinfo{journal}{Phys. Lett.} \textbf{\bibinfo{volume}{B 429}},
  \bibinfo{pages}{263} (\bibinfo{year}{1998}), \eprint{hep-ph/9803315}.

\bibitem[{\citenamefont{Arkani-Hamed et~al.}(2000)\citenamefont{Arkani-Hamed,
  Dimopoulos, Kaloper, and Sundrum}}]{Cosmological2000}
\bibinfo{author}{\bibfnamefont{N.}~\bibnamefont{Arkani-Hamed}},
  \bibinfo{author}{\bibfnamefont{S.}~\bibnamefont{Dimopoulos}},
  \bibinfo{author}{\bibfnamefont{N.}~\bibnamefont{Kaloper}}, \bibnamefont{and}
  \bibinfo{author}{\bibfnamefont{R.}~\bibnamefont{Sundrum}},
  \bibinfo{journal}{Phys. Lett.} \textbf{\bibinfo{volume}{B 480}},
  \bibinfo{pages}{193} (\bibinfo{year}{2000}), \eprint{hep-th/0001197}.

\bibitem[{\citenamefont{Kim et~al.}(2001)\citenamefont{Kim, Kyae, and
  Lee}}]{PhysRevLett.86.4223}
\bibinfo{author}{\bibfnamefont{J.~E.} \bibnamefont{Kim}},
  \bibinfo{author}{\bibfnamefont{B.}~\bibnamefont{Kyae}}, \bibnamefont{and}
  \bibinfo{author}{\bibfnamefont{H.~M.} \bibnamefont{Lee}},
  \bibinfo{journal}{Phys. Rev. Lett.} \textbf{\bibinfo{volume}{86}},
  \bibinfo{pages}{4223} (\bibinfo{year}{2001}), \eprint{hep-th/0011118}.

\bibitem[{\citenamefont{Dey et~al.}(2009)\citenamefont{Dey, Mukhopadhyaya, and
  SenGupta}}]{coscon2009}
\bibinfo{author}{\bibfnamefont{P.}~\bibnamefont{Dey}},
  \bibinfo{author}{\bibfnamefont{B.}~\bibnamefont{Mukhopadhyaya}},
  \bibnamefont{and} \bibinfo{author}{\bibfnamefont{S.}~\bibnamefont{SenGupta}},
  \bibinfo{journal}{Phys. Rev.} \textbf{\bibinfo{volume}{D 80}},
  \bibinfo{pages}{055029} (\bibinfo{year}{2009}), \eprint{0904.1970}.

\bibitem[{\citenamefont{Neupane}(2011)}]{Neupane2010ey}
\bibinfo{author}{\bibfnamefont{I.~P.} \bibnamefont{Neupane}},
  \bibinfo{journal}{Phys. Rev.} \textbf{\bibinfo{volume}{D 83}},
  \bibinfo{pages}{086004} (\bibinfo{year}{2011}), \eprint{1011.6357}.

\bibitem[{\citenamefont{Haghani et~al.}(2012)\citenamefont{Haghani, Sepangi,
  and Shahidi}}]{BraneFR2012}
\bibinfo{author}{\bibfnamefont{Z.}~\bibnamefont{Haghani}},
  \bibinfo{author}{\bibfnamefont{H.~R.} \bibnamefont{Sepangi}},
  \bibnamefont{and} \bibinfo{author}{\bibfnamefont{S.}~\bibnamefont{Shahidi}},
  \bibinfo{journal}{JCAP} \textbf{\bibinfo{volume}{1202}}, \bibinfo{pages}{031}
  (\bibinfo{year}{2012}), \eprint{1201.6448}.

\bibitem[{\citenamefont{George et~al.}(2009)\citenamefont{George, Trodden, and
  Volkas}}]{ThickBrane20093}
\bibinfo{author}{\bibfnamefont{D.~P.} \bibnamefont{George}},
  \bibinfo{author}{\bibfnamefont{M.}~\bibnamefont{Trodden}}, \bibnamefont{and}
  \bibinfo{author}{\bibfnamefont{R.~R.} \bibnamefont{Volkas}},
  \bibinfo{journal}{JHEP} \textbf{\bibinfo{volume}{0902}}, \bibinfo{pages}{035}
  (\bibinfo{year}{2009}), \eprint{0810.3746}.

\bibitem[{\citenamefont{O.~DeWolfe and Karch}(2000)}]{ThickBraneDewolfe}
\bibinfo{author}{\bibfnamefont{S.~G.} \bibnamefont{O.~DeWolfe},
  \bibfnamefont{D.Z.~Freedman}} \bibnamefont{and}
  \bibinfo{author}{\bibfnamefont{A.}~\bibnamefont{Karch}},
  \bibinfo{journal}{Phys. Rev.} \textbf{\bibinfo{volume}{D 62}},
  \bibinfo{pages}{046008} (\bibinfo{year}{2000}), \eprint{hep-th/9909134}.

\bibitem[{\citenamefont{Gregory et~al.}(2000)\citenamefont{Gregory, Rubakov,
  and Sibiryakov}}]{Gregory:2000jc}
\bibinfo{author}{\bibfnamefont{R.}~\bibnamefont{Gregory}},
  \bibinfo{author}{\bibfnamefont{V.}~\bibnamefont{Rubakov}}, \bibnamefont{and}
  \bibinfo{author}{\bibfnamefont{S.~M.} \bibnamefont{Sibiryakov}},
  \bibinfo{journal}{Phys. Rev. Lett.} \textbf{\bibinfo{volume}{84}},
  \bibinfo{pages}{5928} (\bibinfo{year}{2000}), \eprint{hep-th/0002072}.

\bibitem[{\citenamefont{Kaloper et~al.}(2000)\citenamefont{Kaloper,
  March-Russell, Starkman, and Trodden}}]{Kaloper2000jb}
\bibinfo{author}{\bibfnamefont{N.}~\bibnamefont{Kaloper}},
  \bibinfo{author}{\bibfnamefont{J.}~\bibnamefont{March-Russell}},
  \bibinfo{author}{\bibfnamefont{G.~D.} \bibnamefont{Starkman}},
  \bibnamefont{and} \bibinfo{author}{\bibfnamefont{M.}~\bibnamefont{Trodden}},
  \bibinfo{journal}{Phys. Rev. Lett.} \textbf{\bibinfo{volume}{85}},
  \bibinfo{pages}{928} (\bibinfo{year}{2000}), \eprint{hep-ph/0002001}.

\bibitem[{\citenamefont{Wang}(2002)}]{PhysRevD.66.024024}
\bibinfo{author}{\bibfnamefont{A.}~\bibnamefont{Wang}}, \bibinfo{journal}{Phys.
  Rev.} \textbf{\bibinfo{volume}{D 66}}, \bibinfo{pages}{024024}
  (\bibinfo{year}{2002}).

\bibitem[{\citenamefont{Kobayashi et~al.}(2002)\citenamefont{Kobayashi, Koyama,
  and Soda}}]{ThickBrane2002}
\bibinfo{author}{\bibfnamefont{S.}~\bibnamefont{Kobayashi}},
  \bibinfo{author}{\bibfnamefont{K.}~\bibnamefont{Koyama}}, \bibnamefont{and}
  \bibinfo{author}{\bibfnamefont{J.}~\bibnamefont{Soda}},
  \bibinfo{journal}{Phys. Rev.} \textbf{\bibinfo{volume}{D 65}},
  \bibinfo{pages}{064014} (\bibinfo{year}{2002}), \eprint{hep-th/0107025}.

\bibitem[{\citenamefont{Melfo et~al.}(2003)\citenamefont{Melfo, Pantoja, and
  Skirzewski}}]{ThickBrane2003}
\bibinfo{author}{\bibfnamefont{A.}~\bibnamefont{Melfo}},
  \bibinfo{author}{\bibfnamefont{N.}~\bibnamefont{Pantoja}}, \bibnamefont{and}
  \bibinfo{author}{\bibfnamefont{A.}~\bibnamefont{Skirzewski}},
  \bibinfo{journal}{Phys. Rev.} \textbf{\bibinfo{volume}{D 67}},
  \bibinfo{pages}{105003} (\bibinfo{year}{2003}), \eprint{gr-qc/0211081}.

\bibitem[{\citenamefont{Bazeia and Gomes}(2004)}]{Blochbrane2004}
\bibinfo{author}{\bibfnamefont{D.}~\bibnamefont{Bazeia}} \bibnamefont{and}
  \bibinfo{author}{\bibfnamefont{A.}~\bibnamefont{Gomes}},
  \bibinfo{journal}{JHEP} \textbf{\bibinfo{volume}{0405}}, \bibinfo{pages}{012}
  (\bibinfo{year}{2004}), \eprint{hep-th/0403141}.

\bibitem[{\citenamefont{Bazeia et~al.}(2006)\citenamefont{Bazeia, Brito, and
  Losano}}]{ThickBraneBazeia2006}
\bibinfo{author}{\bibfnamefont{D.}~\bibnamefont{Bazeia}},
  \bibinfo{author}{\bibfnamefont{F.}~\bibnamefont{Brito}}, \bibnamefont{and}
  \bibinfo{author}{\bibfnamefont{L.}~\bibnamefont{Losano}},
  \bibinfo{journal}{JHEP} \textbf{\bibinfo{volume}{0611}}, \bibinfo{pages}{064}
  (\bibinfo{year}{2006}), \eprint{hep-th/0610233}.

\bibitem[{\citenamefont{Cardoso et~al.}(2007)\citenamefont{Cardoso, Koyama,
  Mennim, Seahra, and Wands}}]{ThickBrane20071}
\bibinfo{author}{\bibfnamefont{A.}~\bibnamefont{Cardoso}},
  \bibinfo{author}{\bibfnamefont{K.}~\bibnamefont{Koyama}},
  \bibinfo{author}{\bibfnamefont{A.}~\bibnamefont{Mennim}},
  \bibinfo{author}{\bibfnamefont{S.~S.} \bibnamefont{Seahra}},
  \bibnamefont{and} \bibinfo{author}{\bibfnamefont{D.}~\bibnamefont{Wands}},
  \bibinfo{journal}{Phys. Rev.} \textbf{\bibinfo{volume}{D 75}},
  \bibinfo{pages}{084002} (\bibinfo{year}{2007}), \eprint{hep-th/0612202}.

\bibitem[{\citenamefont{V.~Dzhunushaliev and
  Aguilar-Rudametkin}(2008)}]{ThickBraneDzhunshaliev2008}
\bibinfo{author}{\bibfnamefont{D.~S.} \bibnamefont{V.~Dzhunushaliev},
  \bibfnamefont{V.~Folomeev}} \bibnamefont{and}
  \bibinfo{author}{\bibfnamefont{S.}~\bibnamefont{Aguilar-Rudametkin}},
  \bibinfo{journal}{Phys. Rev.} \textbf{\bibinfo{volume}{D 77}},
  \bibinfo{pages}{044006} (\bibinfo{year}{2008}), \eprint{hep-th/0703043}.

\bibitem[{\citenamefont{Liu et~al.}(2010)\citenamefont{Liu, Yang, and
  Zhong}}]{KeYang2009Weyl}
\bibinfo{author}{\bibfnamefont{Y.-X.} \bibnamefont{Liu}},
  \bibinfo{author}{\bibfnamefont{K.}~\bibnamefont{Yang}}, \bibnamefont{and}
  \bibinfo{author}{\bibfnamefont{Y.}~\bibnamefont{Zhong}},
  \bibinfo{journal}{JHEP} \textbf{\bibinfo{volume}{1010}}, \bibinfo{pages}{069}
  (\bibinfo{year}{2010}), \eprint{0911.0269}.

\bibitem[{\citenamefont{Liu et~al.}(2011)\citenamefont{Liu, Zhong, Zhao, and
  Li}}]{ThickBraneZhongYuan2011JHEP}
\bibinfo{author}{\bibfnamefont{Y.-X.} \bibnamefont{Liu}},
  \bibinfo{author}{\bibfnamefont{Y.}~\bibnamefont{Zhong}},
  \bibinfo{author}{\bibfnamefont{Z.-H.} \bibnamefont{Zhao}}, \bibnamefont{and}
  \bibinfo{author}{\bibfnamefont{H.-T.} \bibnamefont{Li}},
  \bibinfo{journal}{JHEP} \textbf{\bibinfo{volume}{1106}}, \bibinfo{pages}{135}
  (\bibinfo{year}{2011}), \eprint{1104.3188}.

\bibitem[{\citenamefont{Liu et~al.}(2012)\citenamefont{Liu, Yang, Guo, and
  Zhong}}]{Liu:2012rc}
\bibinfo{author}{\bibfnamefont{Y.-X.} \bibnamefont{Liu}},
  \bibinfo{author}{\bibfnamefont{K.}~\bibnamefont{Yang}},
  \bibinfo{author}{\bibfnamefont{H.}~\bibnamefont{Guo}}, \bibnamefont{and}
  \bibinfo{author}{\bibfnamefont{Y.}~\bibnamefont{Zhong}},
  \bibinfo{journal}{Phys. Rev.} \textbf{\bibinfo{volume}{D 85}},
  \bibinfo{pages}{124053} (\bibinfo{year}{2012}), \eprint{1203.2349}.

\bibitem[{\citenamefont{Liu et~al.}(2013)\citenamefont{Liu, Wang, Wu, and
  Zhong}}]{Liu:2012mia}
\bibinfo{author}{\bibfnamefont{Y.-X.} \bibnamefont{Liu}},
  \bibinfo{author}{\bibfnamefont{Y.-Q.} \bibnamefont{Wang}},
  \bibinfo{author}{\bibfnamefont{S.-F.} \bibnamefont{Wu}}, \bibnamefont{and}
  \bibinfo{author}{\bibfnamefont{Y.}~\bibnamefont{Zhong}},
  \bibinfo{journal}{Phys. Rev.} \textbf{\bibinfo{volume}{D 88}},
  \bibinfo{pages}{104033} (\bibinfo{year}{2013}), \eprint{1201.5922}.

\bibitem[{\citenamefont{Bazeia et~al.}(2014)\citenamefont{Bazeia, Lobao,
  Menezes, Petrov, and da~Silva}}]{Bazeia:2013uva}
\bibinfo{author}{\bibfnamefont{D.}~\bibnamefont{Bazeia}},
  \bibinfo{author}{\bibfnamefont{A.~J.} \bibnamefont{Lobao},
  \bibfnamefont{A.o}},
  \bibinfo{author}{\bibfnamefont{R.}~\bibnamefont{Menezes}},
  \bibinfo{author}{\bibfnamefont{A.~Y.} \bibnamefont{Petrov}},
  \bibnamefont{and} \bibinfo{author}{\bibfnamefont{A.}~\bibnamefont{da~Silva}},
  \bibinfo{journal}{Phys. Lett.} \textbf{\bibinfo{volume}{B 729}},
  \bibinfo{pages}{127} (\bibinfo{year}{2014}), \eprint{1311.6294}.

\bibitem[{\citenamefont{Das et~al.}(2014)\citenamefont{Das, Mathews, Ravindran,
  and Seth}}]{Das:2014tva}
\bibinfo{author}{\bibfnamefont{G.}~\bibnamefont{Das}},
  \bibinfo{author}{\bibfnamefont{P.}~\bibnamefont{Mathews}},
  \bibinfo{author}{\bibfnamefont{V.}~\bibnamefont{Ravindran}},
  \bibnamefont{and} \bibinfo{author}{\bibfnamefont{S.}~\bibnamefont{Seth}},
  \bibinfo{journal}{JHEP} \textbf{\bibinfo{volume}{1410}}, \bibinfo{pages}{188}
  (\bibinfo{year}{2014}), \eprint{1408.3970}.

\bibitem[{\citenamefont{Grossman and Neubert}(2000)}]{Grossman:1999ra}
\bibinfo{author}{\bibfnamefont{Y.}~\bibnamefont{Grossman}} \bibnamefont{and}
  \bibinfo{author}{\bibfnamefont{M.}~\bibnamefont{Neubert}},
  \bibinfo{journal}{Phys. Lett.} \textbf{\bibinfo{volume}{B 474}},
  \bibinfo{pages}{361} (\bibinfo{year}{2000}), \eprint{hep-ph/9912408}.

\bibitem[{\citenamefont{Bajc and Gabadadze}(2000)}]{Bajc:1999mh}
\bibinfo{author}{\bibfnamefont{B.}~\bibnamefont{Bajc}} \bibnamefont{and}
  \bibinfo{author}{\bibfnamefont{G.}~\bibnamefont{Gabadadze}},
  \bibinfo{journal}{Phys. Lett.} \textbf{\bibinfo{volume}{B 474}},
  \bibinfo{pages}{282} (\bibinfo{year}{2000}), \eprint{hep-th/9912232}.

\bibitem[{\citenamefont{Gremm}(2000)}]{Gremm:1999pj}
\bibinfo{author}{\bibfnamefont{M.}~\bibnamefont{Gremm}},
  \bibinfo{journal}{Phys. Lett.} \textbf{\bibinfo{volume}{B 478}},
  \bibinfo{pages}{434} (\bibinfo{year}{2000}), \eprint{hep-th/9912060}.

\bibitem[{\citenamefont{Chang et~al.}(2000)\citenamefont{Chang, Hisano, Nakano,
  Okada, and Yamaguchi}}]{Chang1999nh}
\bibinfo{author}{\bibfnamefont{S.}~\bibnamefont{Chang}},
  \bibinfo{author}{\bibfnamefont{J.}~\bibnamefont{Hisano}},
  \bibinfo{author}{\bibfnamefont{H.}~\bibnamefont{Nakano}},
  \bibinfo{author}{\bibfnamefont{N.}~\bibnamefont{Okada}}, \bibnamefont{and}
  \bibinfo{author}{\bibfnamefont{M.}~\bibnamefont{Yamaguchi}},
  \bibinfo{journal}{Phys. Rev.} \textbf{\bibinfo{volume}{D 62}},
  \bibinfo{pages}{084025} (\bibinfo{year}{2000}), \eprint{hep-ph/9912498}.

\bibitem[{\citenamefont{Randjbar-Daemi and
  Shaposhnikov}(2000)}]{RandjbarDaemi:2000cr}
\bibinfo{author}{\bibfnamefont{S.}~\bibnamefont{Randjbar-Daemi}}
  \bibnamefont{and} \bibinfo{author}{\bibfnamefont{M.~E.}
  \bibnamefont{Shaposhnikov}}, \bibinfo{journal}{Phys. Lett.}
  \textbf{\bibinfo{volume}{B 492}}, \bibinfo{pages}{361}
  (\bibinfo{year}{2000}), \eprint{hep-th/0008079}.

\bibitem[{\citenamefont{Kehagias and Tamvakis}(2001)}]{Kehagias:2000au}
\bibinfo{author}{\bibfnamefont{A.}~\bibnamefont{Kehagias}} \bibnamefont{and}
  \bibinfo{author}{\bibfnamefont{K.}~\bibnamefont{Tamvakis}},
  \bibinfo{journal}{Phys. Lett.} \textbf{\bibinfo{volume}{B 504}},
  \bibinfo{pages}{38} (\bibinfo{year}{2001}), \eprint{hep-th/0010112}.

\bibitem[{\citenamefont{Duff and Liu}(2001)}]{Duff2000se}
\bibinfo{author}{\bibfnamefont{M.}~\bibnamefont{Duff}} \bibnamefont{and}
  \bibinfo{author}{\bibfnamefont{J.~T.} \bibnamefont{Liu}},
  \bibinfo{journal}{Phys. Lett.} \textbf{\bibinfo{volume}{B 508}},
  \bibinfo{pages}{381} (\bibinfo{year}{2001}), \eprint{hep-th/0010171}.

\bibitem[{\citenamefont{Oda}(2001)}]{Oda2001}
\bibinfo{author}{\bibfnamefont{I.}~\bibnamefont{Oda}}, \bibinfo{journal}{Phys.
  Lett.} \textbf{\bibinfo{volume}{B 508}}, \bibinfo{pages}{96}
  (\bibinfo{year}{2001}), \eprint{hep-th/0012013}.

\bibitem[{\citenamefont{Ringeval et~al.}(2002)\citenamefont{Ringeval, Peter,
  and Uzan}}]{Ringeval:2001cq}
\bibinfo{author}{\bibfnamefont{C.}~\bibnamefont{Ringeval}},
  \bibinfo{author}{\bibfnamefont{P.}~\bibnamefont{Peter}}, \bibnamefont{and}
  \bibinfo{author}{\bibfnamefont{J.-P.} \bibnamefont{Uzan}},
  \bibinfo{journal}{Phys. Rev.} \textbf{\bibinfo{volume}{D 65}},
  \bibinfo{pages}{044016} (\bibinfo{year}{2002}), \eprint{hep-th/0109194}.

\bibitem[{\citenamefont{Ichinose}(2002)}]{Ichinose:2002kg}
\bibinfo{author}{\bibfnamefont{S.}~\bibnamefont{Ichinose}},
  \bibinfo{journal}{Phys. Rev.} \textbf{\bibinfo{volume}{D 66}},
  \bibinfo{pages}{104015} (\bibinfo{year}{2002}), \eprint{hep-th/0206187}.

\bibitem[{\citenamefont{Koley and Kar}(2005)}]{Koley:2004at}
\bibinfo{author}{\bibfnamefont{R.}~\bibnamefont{Koley}} \bibnamefont{and}
  \bibinfo{author}{\bibfnamefont{S.}~\bibnamefont{Kar}},
  \bibinfo{journal}{Class. Quant. Grav.} \textbf{\bibinfo{volume}{22}},
  \bibinfo{pages}{753} (\bibinfo{year}{2005}), \eprint{hep-th/0407158}.

\bibitem[{\citenamefont{Davies et~al.}(2008)\citenamefont{Davies, George, and
  Volkas}}]{Davies:2007xr}
\bibinfo{author}{\bibfnamefont{R.}~\bibnamefont{Davies}},
  \bibinfo{author}{\bibfnamefont{D.~P.} \bibnamefont{George}},
  \bibnamefont{and} \bibinfo{author}{\bibfnamefont{R.~R.}
  \bibnamefont{Volkas}}, \bibinfo{journal}{Phys. Rev.}
  \textbf{\bibinfo{volume}{D 77}}, \bibinfo{pages}{124038}
  (\bibinfo{year}{2008}), \eprint{0705.1584}.

\bibitem[{\citenamefont{Liu et~al.}(2008)\citenamefont{Liu, Zhang, Wei, and
  Duan}}]{Liu2008WeylPT}
\bibinfo{author}{\bibfnamefont{Y.-X.} \bibnamefont{Liu}},
  \bibinfo{author}{\bibfnamefont{L.-D.} \bibnamefont{Zhang}},
  \bibinfo{author}{\bibfnamefont{S.-W.} \bibnamefont{Wei}}, \bibnamefont{and}
  \bibinfo{author}{\bibfnamefont{Y.-S.} \bibnamefont{Duan}},
  \bibinfo{journal}{JHEP} \textbf{\bibinfo{volume}{0808}}, \bibinfo{pages}{041}
  (\bibinfo{year}{2008}), \eprint{0803.0098}.

\bibitem[{\citenamefont{Archer and Huber}(2011)}]{Archer2011}
\bibinfo{author}{\bibfnamefont{P.~R.} \bibnamefont{Archer}} \bibnamefont{and}
  \bibinfo{author}{\bibfnamefont{S.~J.} \bibnamefont{Huber}},
  \bibinfo{journal}{JHEP} \textbf{\bibinfo{volume}{1103}}, \bibinfo{pages}{018}
  (\bibinfo{year}{2011}), \eprint{1010.3588}.

\bibitem[{\citenamefont{Jones et~al.}(2013)\citenamefont{Jones, Munoz,
  Singleton, and Triyanta}}]{Jones:2013ofa}
\bibinfo{author}{\bibfnamefont{P.}~\bibnamefont{Jones}},
  \bibinfo{author}{\bibfnamefont{G.}~\bibnamefont{Munoz}},
  \bibinfo{author}{\bibfnamefont{D.}~\bibnamefont{Singleton}},
  \bibnamefont{and} \bibinfo{author}{\bibnamefont{Triyanta}},
  \bibinfo{journal}{Phys. Rev.} \textbf{\bibinfo{volume}{D 88}},
  \bibinfo{pages}{025048} (\bibinfo{year}{2013}), \eprint{1307.3599}.

\bibitem[{\citenamefont{Xie et~al.}(2013)\citenamefont{Xie, Yang, and
  Zhao}}]{Xie2013rka}
\bibinfo{author}{\bibfnamefont{Q.-Y.} \bibnamefont{Xie}},
  \bibinfo{author}{\bibfnamefont{J.}~\bibnamefont{Yang}}, \bibnamefont{and}
  \bibinfo{author}{\bibfnamefont{L.}~\bibnamefont{Zhao}},
  \bibinfo{journal}{Phys. Rev.} \textbf{\bibinfo{volume}{D 88}},
  \bibinfo{pages}{105014} (\bibinfo{year}{2013}), \eprint{1310.4585}.

\bibitem[{\citenamefont{Cembranos et~al.}(2013)\citenamefont{Cembranos,
  Delgado, and Dobado}}]{Cembranos:2013qja}
\bibinfo{author}{\bibfnamefont{J.~A.~R.} \bibnamefont{Cembranos}},
  \bibinfo{author}{\bibfnamefont{R.~L.} \bibnamefont{Delgado}},
  \bibnamefont{and} \bibinfo{author}{\bibfnamefont{A.}~\bibnamefont{Dobado}},
  \bibinfo{journal}{Phys. Rev.} \textbf{\bibinfo{volume}{D 88}},
  \bibinfo{pages}{075021} (\bibinfo{year}{2013}), \eprint{1306.4900}.

\bibitem[{\citenamefont{Costa et~al.}(2013)\citenamefont{Costa, Silva, and
  Almeida}}]{Costa:2013eua}
\bibinfo{author}{\bibfnamefont{F.}~\bibnamefont{Costa}},
  \bibinfo{author}{\bibfnamefont{J.}~\bibnamefont{Silva}}, \bibnamefont{and}
  \bibinfo{author}{\bibfnamefont{C.}~\bibnamefont{Almeida}},
  \bibinfo{journal}{Phys. Rev.} \textbf{\bibinfo{volume}{D 87}},
  \bibinfo{pages}{125010} (\bibinfo{year}{2013}), \eprint{1304.7825}.

\bibitem[{\citenamefont{Zhao et~al.}(2014)\citenamefont{Zhao, Liu, and
  Zhong}}]{Zhao2014gka}
\bibinfo{author}{\bibfnamefont{Z.-H.} \bibnamefont{Zhao}},
  \bibinfo{author}{\bibfnamefont{Y.-X.} \bibnamefont{Liu}}, \bibnamefont{and}
  \bibinfo{author}{\bibfnamefont{Y.}~\bibnamefont{Zhong}},
  \bibinfo{journal}{Phys. Rev.} \textbf{\bibinfo{volume}{D 90}},
  \bibinfo{pages}{045031} (\bibinfo{year}{2014}), \eprint{1402.6480}.

\bibitem[{\citenamefont{Carrillo-Gonzalez
  et~al.}(2014)\citenamefont{Carrillo-Gonzalez, German, Herrera-Aguilar,
  Hidalgo, and Malagon-Morejon}}]{Carrillo-Gonzalez:2014dka}
\bibinfo{author}{\bibfnamefont{M.}~\bibnamefont{Carrillo-Gonzalez}},
  \bibinfo{author}{\bibfnamefont{G.}~\bibnamefont{German}},
  \bibinfo{author}{\bibfnamefont{A.}~\bibnamefont{Herrera-Aguilar}},
  \bibinfo{author}{\bibfnamefont{J.~C.} \bibnamefont{Hidalgo}},
  \bibnamefont{and}
  \bibinfo{author}{\bibfnamefont{D.}~\bibnamefont{Malagon-Morejon}}
  (\bibinfo{year}{2014}), \eprint{1409.5926}.

\bibitem[{\citenamefont{Kulaxizi and Rahman}(2014)}]{Kulaxizi:2014yxa}
\bibinfo{author}{\bibfnamefont{M.}~\bibnamefont{Kulaxizi}} \bibnamefont{and}
  \bibinfo{author}{\bibfnamefont{R.}~\bibnamefont{Rahman}},
  \bibinfo{journal}{JHEP} \textbf{\bibinfo{volume}{1410}}, \bibinfo{pages}{193}
  (\bibinfo{year}{2014}), \eprint{1409.1942}.

\bibitem[{\citenamefont{Koivisto and Nunes}(2013)}]{PhysRevD.88.123512}
\bibinfo{author}{\bibfnamefont{T.~S.} \bibnamefont{Koivisto}} \bibnamefont{and}
  \bibinfo{author}{\bibfnamefont{N.~J.} \bibnamefont{Nunes}},
  \bibinfo{journal}{Phys. Rev.} \textbf{\bibinfo{volume}{D 88}},
  \bibinfo{pages}{123512} (\bibinfo{year}{2013}),
  \urlprefix\url{http://link.aps.org/doi/10.1103/PhysRevD.88.123512}.

\bibitem[{\citenamefont{Rindani and Sivakumar}(1985)}]{PhysRevD.32.3238}
\bibinfo{author}{\bibfnamefont{S.~D.} \bibnamefont{Rindani}} \bibnamefont{and}
  \bibinfo{author}{\bibfnamefont{M.}~\bibnamefont{Sivakumar}},
  \bibinfo{journal}{Phys. Rev.} \textbf{\bibinfo{volume}{D 32}},
  \bibinfo{pages}{3238} (\bibinfo{year}{1985}).

\bibitem[{\citenamefont{Sivakumar}(1988)}]{PhysRevD.37.1690}
\bibinfo{author}{\bibfnamefont{M.}~\bibnamefont{Sivakumar}},
  \bibinfo{journal}{Phys. Rev.} \textbf{\bibinfo{volume}{D 37}},
  \bibinfo{pages}{1690} (\bibinfo{year}{1988}).

\bibitem[{\citenamefont{Mukhopadhyaya et~al.}(2002)\citenamefont{Mukhopadhyaya,
  Sen, and SenGupta}}]{PRLKR2002}
\bibinfo{author}{\bibfnamefont{B.}~\bibnamefont{Mukhopadhyaya}},
  \bibinfo{author}{\bibfnamefont{S.}~\bibnamefont{Sen}}, \bibnamefont{and}
  \bibinfo{author}{\bibfnamefont{S.}~\bibnamefont{SenGupta}},
  \bibinfo{journal}{Phys. Rev. Lett.} \textbf{\bibinfo{volume}{89}},
  \bibinfo{pages}{121101} (\bibinfo{year}{2002}), \eprint{hep-th/0204242}.

\bibitem[{\citenamefont{Mukhopadhyaya et~al.}(2007)\citenamefont{Mukhopadhyaya,
  Sen, and SenGupta}}]{QformRS}
\bibinfo{author}{\bibfnamefont{B.}~\bibnamefont{Mukhopadhyaya}},
  \bibinfo{author}{\bibfnamefont{S.}~\bibnamefont{Sen}}, \bibnamefont{and}
  \bibinfo{author}{\bibfnamefont{S.}~\bibnamefont{SenGupta}},
  \bibinfo{journal}{Phys. Rev.} \textbf{\bibinfo{volume}{D 76}},
  \bibinfo{pages}{121501} (\bibinfo{year}{2007}), \eprint{0709.3428}.

\bibitem[{\citenamefont{Alencar
  et~al.}(2010{\natexlab{a}})\citenamefont{Alencar, Tahim, Landim, Muniz, and
  Costa~Filho}}]{qRSdilation-eprintv2}
\bibinfo{author}{\bibfnamefont{G.}~\bibnamefont{Alencar}},
  \bibinfo{author}{\bibfnamefont{M.}~\bibnamefont{Tahim}},
  \bibinfo{author}{\bibfnamefont{R.}~\bibnamefont{Landim}},
  \bibinfo{author}{\bibfnamefont{C.}~\bibnamefont{Muniz}}, \bibnamefont{and}
  \bibinfo{author}{\bibfnamefont{R.}~\bibnamefont{Costa~Filho}},
  \bibinfo{journal}{Phys. Rev.} \textbf{\bibinfo{volume}{D 82}},
  \bibinfo{pages}{104053} (\bibinfo{year}{2010}{\natexlab{a}}),
  \eprint{1005.1691}.

\bibitem[{\citenamefont{Alencar
  et~al.}(2010{\natexlab{b}})\citenamefont{Alencar, Landim, Tahim, Muniz, and
  Costa~Filho}}]{QformThickRS}
\bibinfo{author}{\bibfnamefont{G.}~\bibnamefont{Alencar}},
  \bibinfo{author}{\bibfnamefont{R.}~\bibnamefont{Landim}},
  \bibinfo{author}{\bibfnamefont{M.}~\bibnamefont{Tahim}},
  \bibinfo{author}{\bibfnamefont{C.}~\bibnamefont{Muniz}}, \bibnamefont{and}
  \bibinfo{author}{\bibfnamefont{R.}~\bibnamefont{Costa~Filho}},
  \bibinfo{journal}{Phys. Lett.} \textbf{\bibinfo{volume}{B 693}},
  \bibinfo{pages}{503} (\bibinfo{year}{2010}{\natexlab{b}}),
  \eprint{1008.0678}.

\bibitem[{\citenamefont{Youm}(2001)}]{Youm:2001dt}
\bibinfo{author}{\bibfnamefont{D.}~\bibnamefont{Youm}}, \bibinfo{journal}{Phys.
  Rev.} \textbf{\bibinfo{volume}{D 64}}, \bibinfo{pages}{127501}
  (\bibinfo{year}{2001}), \eprint{hep-th/0106240}.

\bibitem[{\citenamefont{Fu et~al.}(2012)\citenamefont{Fu, Liu, Yang, and
  Wei}}]{Fu2012sa}
\bibinfo{author}{\bibfnamefont{C.-E.} \bibnamefont{Fu}},
  \bibinfo{author}{\bibfnamefont{Y.-X.} \bibnamefont{Liu}},
  \bibinfo{author}{\bibfnamefont{K.}~\bibnamefont{Yang}}, \bibnamefont{and}
  \bibinfo{author}{\bibfnamefont{S.-W.} \bibnamefont{Wei}},
  \bibinfo{journal}{JHEP} \textbf{\bibinfo{volume}{1210}}, \bibinfo{pages}{060}
  (\bibinfo{year}{2012}), \eprint{1207.3152}.

\bibitem[{\citenamefont{Fu et~al.}(2014)\citenamefont{Fu, Liu, Guo, Chen, and
  Zhang}}]{Fu:2013ita}
\bibinfo{author}{\bibfnamefont{C.-E.} \bibnamefont{Fu}},
  \bibinfo{author}{\bibfnamefont{Y.-X.} \bibnamefont{Liu}},
  \bibinfo{author}{\bibfnamefont{H.}~\bibnamefont{Guo}},
  \bibinfo{author}{\bibfnamefont{F.-W.} \bibnamefont{Chen}}, \bibnamefont{and}
  \bibinfo{author}{\bibfnamefont{S.-L.} \bibnamefont{Zhang}},
  \bibinfo{journal}{Phys. Lett.} \textbf{\bibinfo{volume}{B 735}},
  \bibinfo{pages}{7} (\bibinfo{year}{2014}), \eprint{1312.2647}.

\bibitem[{\citenamefont{Ho and Ma}(2013)}]{Ho:2013paa}
\bibinfo{author}{\bibfnamefont{P.-M.} \bibnamefont{Ho}} \bibnamefont{and}
  \bibinfo{author}{\bibfnamefont{C.-T.} \bibnamefont{Ma}},
  \bibinfo{journal}{JHEP} \textbf{\bibinfo{volume}{1305}}, \bibinfo{pages}{056}
  (\bibinfo{year}{2013}), \eprint{1302.6919}.

\bibitem[{\citenamefont{Hinterbichler et~al.}(2014)\citenamefont{Hinterbichler,
  Levin, and Zukowski}}]{Hinterbichler:2013kwa}
\bibinfo{author}{\bibfnamefont{K.}~\bibnamefont{Hinterbichler}},
  \bibinfo{author}{\bibfnamefont{J.}~\bibnamefont{Levin}}, \bibnamefont{and}
  \bibinfo{author}{\bibfnamefont{C.}~\bibnamefont{Zukowski}},
  \bibinfo{journal}{Phys. Rev.} \textbf{\bibinfo{volume}{D 89}},
  \bibinfo{pages}{086007} (\bibinfo{year}{2014}), \eprint{1310.6353}.

\bibitem[{\citenamefont{Jardim et~al.}(2014)\citenamefont{Jardim, Alencar,
  Landim, and Costa~Filho}}]{Jardim2014vba}
\bibinfo{author}{\bibfnamefont{I.}~\bibnamefont{Jardim}},
  \bibinfo{author}{\bibfnamefont{G.}~\bibnamefont{Alencar}},
  \bibinfo{author}{\bibfnamefont{R.}~\bibnamefont{Landim}}, \bibnamefont{and}
  \bibinfo{author}{\bibfnamefont{R.}~\bibnamefont{Costa~Filho}}
  (\bibinfo{year}{2014}), \eprint{1410.6756}.

\bibitem[{\citenamefont{Ho and Ma}(2014)}]{Ho2014una}
\bibinfo{author}{\bibfnamefont{J.-K.} \bibnamefont{Ho}} \bibnamefont{and}
  \bibinfo{author}{\bibfnamefont{C.-T.} \bibnamefont{Ma}}
  (\bibinfo{year}{2014}), \eprint{1410.0972}.

\bibitem[{\citenamefont{Smailagic and Spallucci}(2000)}]{Smailagic2000hr}
\bibinfo{author}{\bibfnamefont{A.}~\bibnamefont{Smailagic}} \bibnamefont{and}
  \bibinfo{author}{\bibfnamefont{E.}~\bibnamefont{Spallucci}},
  \bibinfo{journal}{Phys.Lett.} \textbf{\bibinfo{volume}{B489}},
  \bibinfo{pages}{435} (\bibinfo{year}{2000}), \eprint{hep-th/0008094}.

\end{thebibliography}

\end{document}